\documentclass[structabstract]{aa}  
\usepackage{graphicx}
\usepackage{txfonts}
\usepackage{natbib}
\usepackage{longtable}
\usepackage{lscape} 
\usepackage{graphicx}
\usepackage{url}
\usepackage{amssymb}
\usepackage{deluxetable}

\newlength{\jacobwidth}
\def\jacob#1#2#3{\settowidth{\jacobwidth}{$_{#3}$}
	\left(\frac{d#1}{d#2}\right)_{_{\!\!#3}}\hspace{-\jacobwidth}}

\begin{document}
   \title{Dynamical Analysis of Nearby ClustErs \thanks{Based on observations obtained with MegaPrime/MegaCam, a joint project of CFHT and CEA/DAPNIA, at the Canada-France-Hawaii Telescope (CFHT) which is operated by the National Research Council (NRC) of Canada, the Institut National des Science de l'Univers of the Centre National de la Recherche Scientifique (CNRS) of France, and the University of Hawaii.}}
   \subtitle{Automated astrometry from the ground: precision proper motions over wide field}
   \author{H. Bouy \inst{1}
          \and E. Bertin\inst{2}
          \and E. Moraux\inst{3}
          \and J.-C. Cuillandre\inst{4}
          \and J. Bouvier\inst{3}
          \and D. Barrado\inst{1,5}
          \and E. Solano\inst{1}
          \and A. Bayo\inst{6}
          }

   \institute{Centro de Astrobiolog\'\i a, depto de Astrof\'\i sica, INTA-CSIC, PO BOX 78, E-28691, ESAC Campus, Villanueva de la Ca\~nada, Madrid, Spain\\
         \email{hbouy@cab.inta-csic.es}
         \and 
         Institut d'Astrophysique de Paris, CNRS UMR 7095 and UPMC, 98bis bd Arago, F-75014 Paris, France \\
         \and
         UJF-Grenoble 1/CNRS-INSU, Institut de Plan\'etologie et d'Astrophysique de Grenoble (IPAG), UMR 5274, Grenoble, F-38041, France\\
         \and
         Canada-France-Hawaii Telescope Corporation, 65-1238 Mamalahoa Highway, Kamuela, HI96743, USA \\
         \and
         Calar Alto Observatory, Centro Astron\'omico Hispano Alem\'an, Calle Jes\'us Durb\'an Rem\'on, 04004 Almer\'\i a, Spain\\
         \and
         European Southern Observatory, Alonso de Cordova 3107, Vitacura, Santiago, Chile            }

   \date{Received ; accepted }

% \abstract{}{}{}{}{} 
%  {} token are mandatory
 
  \abstract{The kinematic properties of the different classes of objects in a given association hold important clues about its member’s history, and offer a unique opportunity to test the predictions of the various models of stellar formation and evolution.}
   {DANCe (standing for Dynamical Analysis of Nearby ClustErs) is a survey program aimed at deriving a comprehensive and homogeneous census of the stellar and substellar content of a number of nearby ($<$1~kpc) young ($<$500~Myr) associations. Whenever possible, members will be identified based on their kinematics properties, ensuring little contamination from background and foreground sources. Otherwise, the dynamics of previously confirmed members will be studied using the proper motion measurements. We present here the method used to derive precise proper motion measurements, using the Pleiades cluster as a test bench.  }
   {Combining deep wide field multi-epoch panchromatic images obtained at various obervatories over up to 14~years, we derive accurate proper motions for the sources present in the field of the survey. The datasets cover $\approx$80 square degrees, centered around the Seven Sisters.}
   {Using new tools, we have computed a catalog of 6\,116\,907 unique sources, including proper motion measurements for 3\,577\,478 of them. The catalogue covers the magnitude range between $i=$12$\sim$24~mag, achieving a proper motion accuracy $<$1~mas y$^{-1}$ for sources as faint as $i$=22.5~mag. We estimate that our final accuracy reaches 0.3~mas yr$^{-1}$ in the best cases, depending on magnitude, observing history,  and the presence of reference extragalactic sources for the anchoring onto the ICRS.}
   {}


   \maketitle
%
%________________________________________________________________

\section{Introduction}
The Milky Way galaxy includes large scale structures such as clusters, star forming regions, and OB associations. Understanding the formation, structure, and evolution of these components has been one of the greatest challenges of modern astrophysics. Following the advent of sensitive wide-field instruments over the past two decades, a large number of photometric studies have been performed in stellar associations and clusters \citep[e.g ][]{2000A&A...360..539K, 2004ApJ...610.1064B,2001ApJ...556..830B,2006A&A...450..147L,2008hsf2.book..757T,1992A&A...262..468E}. These surveys not only dramatically improved our knowledge of the luminosity function, but also extended it down to the substellar and planetary mass regimes. They nevertheless suffer from several limitations, making their comparison to theoretical predictions sometimes difficult. Any photometric selection indeed relies on theoretical tracks, and hence age estimates, that are still uncertain at young ages \citep{2009ApJ...702L..27B}. Additionally, the photometric variability inherent to their youth can affect their luminosity and colors, leading to a significant fraction of missed members. Foreground stars and extragalactic sources will in most cases be a major source of contamination. Finally, photometric surveys are not able to disentangle members of neighboring or spatially coincident groups, possibly leading to confusion and erroneous conclusions regarding their origin. Selecting members based on their kinematics offers several advantages: it is completely independant of evolutionary models; it rejects the majority of unrelated foreground and background sources; it is insensitive to variability or flux excess and deficiency related to e.g circumstellar material or accretion; it can disentangle coincident or neighboring associations provided that they have differing mean motions  \citep[e.g $\sigma-$Ori, Lupus, and Upper Scorpius,][]{2006MNRAS.371L...6J, 1996A&A...307..121B,2000A&A...356..541K,2007ApJ...662..413K, 2011A&A...529A.108L}. 

The study of kinematics involves two complementary observational techniques: radial and transverse velocity measurements. Systematic radial velocity surveys over extended ($>$10~deg$^2$) regions of the sky require large amount of telescope time. The most succesful and efficient spectroscopic surveys to date \citep[e.g RAVE, WOCS,  APOGEE, MARVELS, ESO-GAIA, ][]{RAVE,WOCS, APOGEE, MARVELS,2012Msngr.147...25G} are producing libraries including several hundreds of thousands of high quality spectra and radial velocity measurements over areas as large as several thousands square degrees. They are nevertheless still limited to the brightest sources and do not reach the substellar luminosity range. Proper motion measurements are on the other hand much easier to achieve. One in principle only needs two observations separated by a sufficient period of time. A number of nearby associations have mean  proper motions of a few tens of mas yr$^{-1}$ \citep[e.g Pleiades, Hyades, Taurus, Ophiuchus,][]{2005A&A...438.1163K,2006AstL...32..816B} making it easier to measure their members' motion over a few years only. In spite of the tremendous efforts conducted over the past 20~yr, the limited sensitivity of the various large scale kinematics projects has restricted the study to the Solar neighbourhood (e.g Hipparcos) or to the identification of nearby moving groups \citep{2006A&A...460..695T,2008hsf2.book..757T,2004ApJ...613L..65Z,2001ApJ...549L.233Z,2006ApJ...649L.115Z}. Local kinematics has been commonly used to confirm photometrically selected samples \citep[see e.g][]{2001A&A...367..211M,2003A&A...400..891M,2007ApJ...662..413K,2007MNRAS.380..712L,2012MNRAS.422.1495L,2007MNRAS.374..372L,2010MNRAS.tmp.1513G, 2011A&A...529A.108L}, but kinematically selected samples over large areas of young nearby associations are still sorely missing. 

The future {\it Gaia} space mission \citep[][]{2001A&A...369..339P} will provide an exquisite accuracy and complete 6 dimension census of the sky up to G$\approx$15~mag, and a 5 dimension census up to G$\approx$20~mag. Although it represents a tremendous improvement with respect to its predecessor {\it Hipparcos} \citep{1997A&A...323L..49P}, {\it Gaia} will unfortunately not be sensitive enough to study the least massive objects. G$\approx$20~mag indeed corresponds to $\approx$15~M$_{\rm Jup}$ at 150~pc and for an age of 1~Myr \citep{1998A&A...337..403B}, when the mass function is known to extend at least down to 3$\sim$4~M$_{\rm Jup}$ \citep[e.g ][ and references therein]{2011A&A...536A..63B}. Additionally, young stellar clusters and associations are very often deeply embedded and contain bright H~{\sc II} regions. Since it will operate in the visible part of the spectrum, {\it Gaia} will be mostly blind in the regions of heavy extinction and bright nebular emission, where precisely most of the star formation is taking place. 

Recently, \citet{2006A&A...454.1029A} demonstrated that high-precision astrometry could be extracted from wide-field, ground-based CCD images, and studied extended areas around galactic globular clusters \citep{2009A&A...493..959B,2010A&A...517A..34B,2008A&A...484..609Y} using observations obtained with the ESO {\it WFI} wide-field camera. In this paper, we present a similar method designed to automatically process and analyse vast amounts of images (several thousands) originating from multiple instruments and sites and covering large ($>$10~deg$^2$) areas of the sky.

\section{The DANCe Project: \label{dance}}

Taking advantage of the wide field surveys performed in the early 2000, we are performing a comprehensive study of kinematics in a number of nearby ($\lesssim$1~kpc)  associations and clusters. A preliminary list of targets with publicly available archival observations is given in \citet{2011sca..conf..103B}, and we are welcoming suggestions and proposals of collaborative studies for other associations and clusters of particular interest. The initial surveys reached sensitivities well beyond the substellar limit at the age and distance of these associations. Complementing these archival data with new sensitive wide field observations, we are compiling a multi-epoch panchromatic database encompassing large (several tens of square degrees) areas of young nearby associations. This database is used to derive accurate proper motions for all sources with multi-epoch detections. The scientific goals of the DANCe project are twofold:

\begin{itemize}
\item {\it mass function:} when the association mean proper motion allows it, the proper motion and accurate photometric measurements can be used to select members and/or reject contaminants and derive more accurate luminosity (mass) functions. The samples can be used to study the stellar content within each group separately, and to perform a meaningful intercomparison between groups of various ages, structures, metallicities, and densities. 
\item {\it internal dynamics:} the observed velocity distribution of confirmed membersand its dependance on stellar mass, spatial distribution and environment can be compared with advanced N-Body numerical simulations and dynamical evolution models \citep{2001ASPC..243..291A,2009ApJS..185..486P,2012arXiv1205.1508M}. Ultimately, complementary radial velocity measurements can provide a complete picture of the space motions within young associations, below the substellar limit. 
\end{itemize}

\section{Test case: the Pleiades}
Their youth and proximity have made the Pleiades one of the most extensively studied clusters over the past hundred years.
In their recent review of the cluster, \citet{2007ApJS..172..663S} and \citet{2012MNRAS.422.1495L} have compiled an exhaustive list of candidate and confirmed members originating from more than a dozen independent surveys of the Pleiades. The total number of members and candidate members reported in their catalogs adds up to 1471 objects. The relatively large mean proper motion of the group and the vast amounts of images available in public archives makes it an ideal target to develop the large scale data processing and automatic astrometric algorithms presented in this manuscript.

\section{Archival Data \label{archive}}

\begin{figure*}
   \centering
   \includegraphics[width=0.95\textwidth]{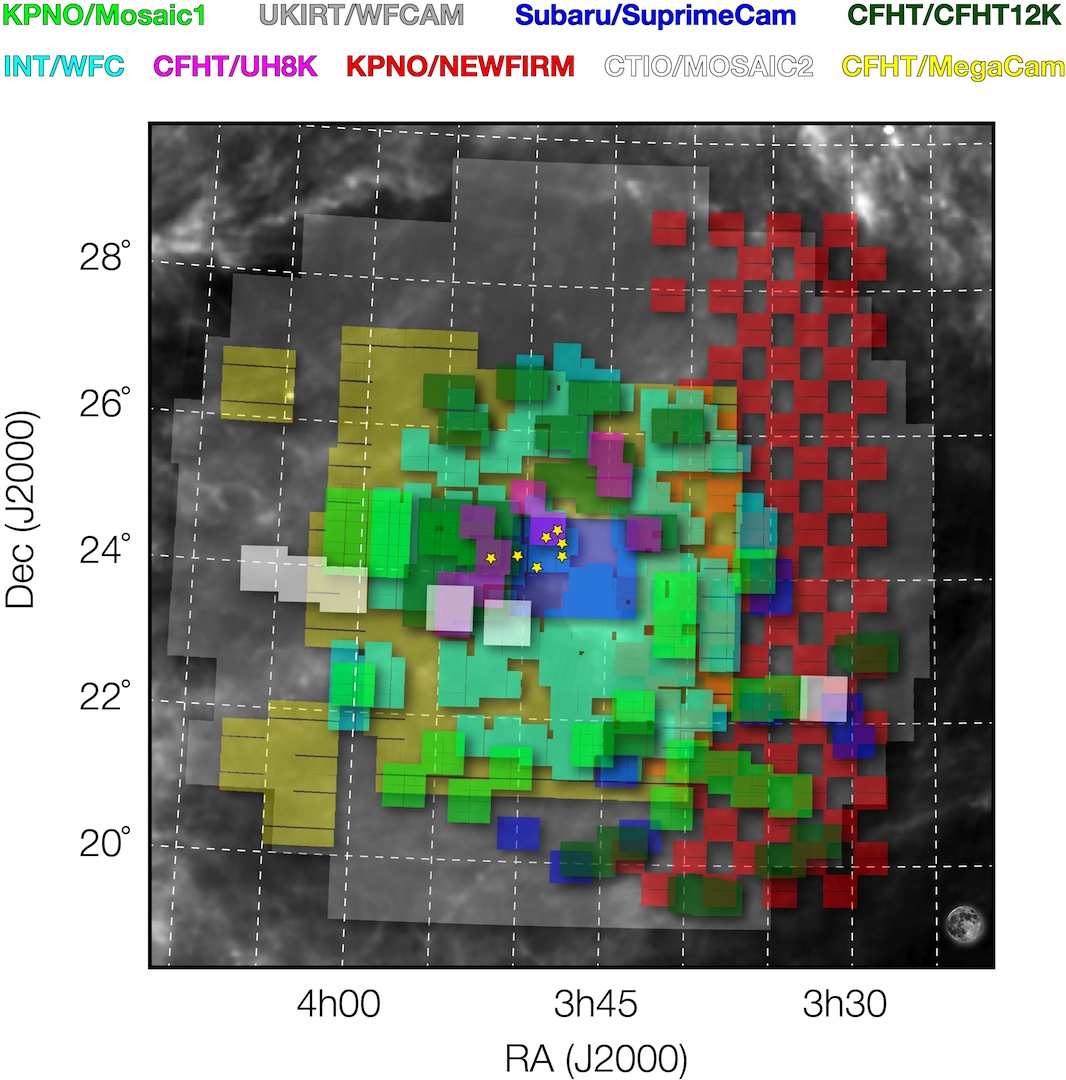}
      \caption{IRAS 100~$\mu$m image of the Pleiades cluster with the various surveys used in this study overplotted. The Seven Sisters are represented with yellow stars. The full Moon is represented in the lower right corner to illustrate the scale. }
         \label{surveys}
   \end{figure*}

\begin{figure*}
   \centering
   \includegraphics[width=0.95\textwidth]{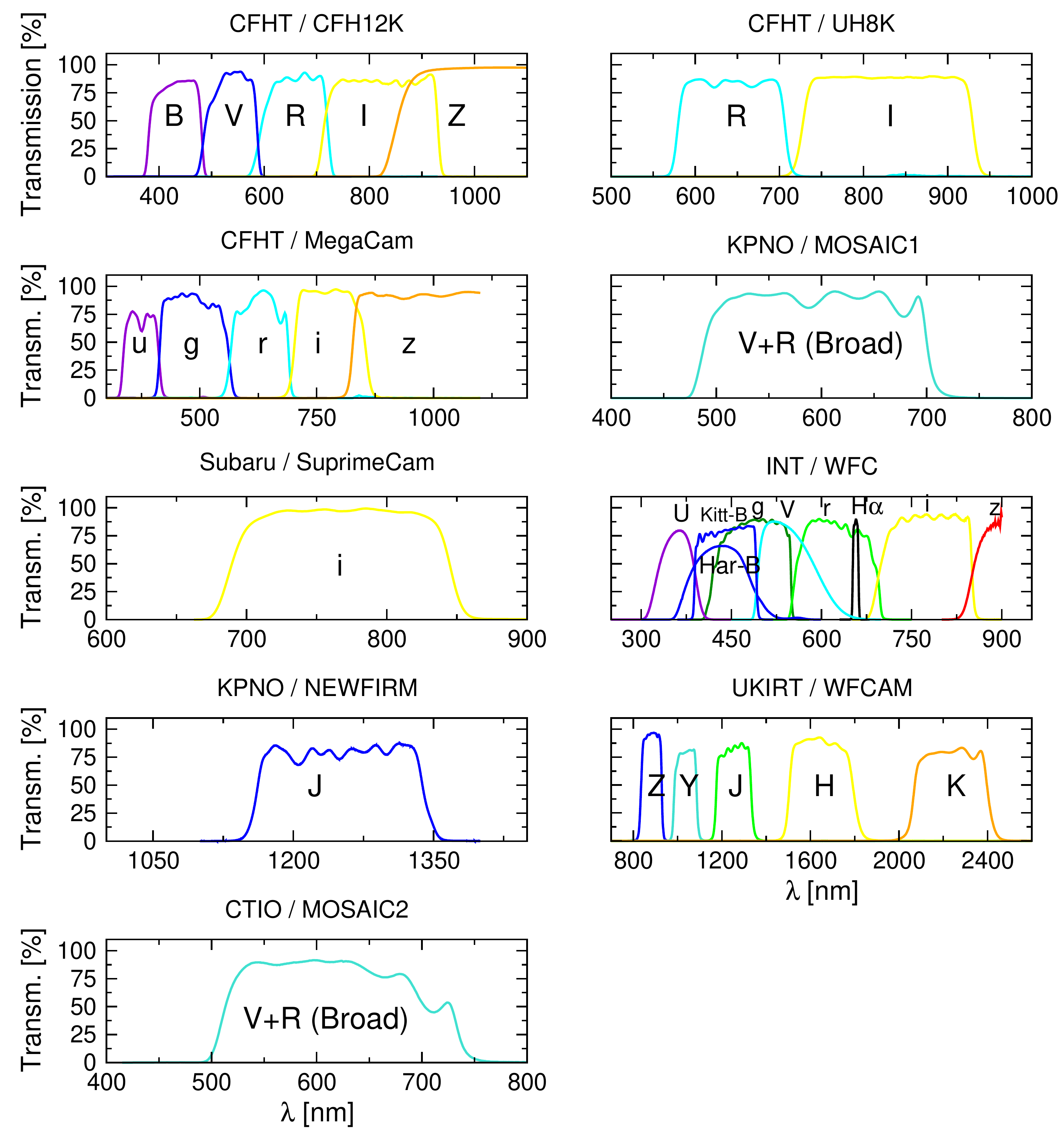}
      \caption{Transmission of the various filters used in this study.  In the $z$-band, the sensitivity is limited by the CCD quantum efficiency, which typically drops at 900$\sim$1000~nm. }
         \label{filters}
   \end{figure*}

In an effort to compile the most complete dataset - both in terms of spatial and time coverage - we searched the {\it Subaru} Telescope, the {\it Isaac Newton Telescope (INT)}, the {\it United Kingdom Infrared Telescope (UKIRT)}, the {\it Cerro Tololo Inter-American Observatory (CTIO, at NOAO)}, the {\it Kitt Peak National Observatory (KPNO, at NOAO)}, the {\it Canada France Hawai'i Telescope (CFHT)} and the {\it European Southern Observatory (ESO)} public archives for wide-field images within a box of 10\degr$\times$10\degr\, centered on the Pleiades. Figure~\ref{surveys} and Table~\ref{table_obs} give an overview of the properties and coverage of the various datasets and instruments. The filter sets used for these observations are described in Fig.~\ref{filters}\footnote{The transmission curves were retrieved from the Spanish VO website http://svo2.cab.inta-csic.es/theory/fps/}. The data were obtained with 9 different instruments at 5 observatories. A summary of their characteristics is given in the following sections. 

\begin{itemize}
\item[$\bullet$] CFH  Telescope:\\
The {\it CFH12K} \citep{2000SPIE.4008.1010C}  and {\it UH8K} \citep{1995AAS...187.7305M} observations of the Pleiades are described in details in
\citet{1998A&A...336..490B,2001A&A...367..211M} and \citet{2003A&A...400..891M}.
The {\it MegaCam} \citep{2003SPIE.4841...72B}  observations are described in Section~\ref{newobs}. The
individual images were processed and calibrated with the recommended
{\it Elixir} system \citep{2004PASP..116..449M}, which includes detrending
(darks, biases, flats and fringe frames), and astrometric registration. Nightly
magnitude zero-points were derived by the CFHT team using standard star fields
\citep{1992AJ....104..340L}.

\item[$\bullet$] Subaru  Telescope: \\
The {\it Suprime-Cam} \citep{2002PASJ...54..833M} images were processed (overscan subtraction,
flat-fielding, and masking of vignetted areas) using the recommended SDFRED1
package \citep{2004ApJ...611..660O,2002AJ....123...66Y} and the relevant
calibration frames obtained the same night. The photometric conditions on
Mauna Kea were poor during these observations, as described in the Skyprobe
database \citep{2004sdab.conf..287C}. We therefore did not attempt to calibrate
the corresponding photometry.

\item[$\bullet$] INT Telescope:\\
We retrieved the detrended individual  {\it Wide Field Camera} \citep[{\it WFC},][]{1998IEEES..16...20I} images from the ING
public archive. About 88\% of these observations were obtained under photometric
ambient conditions, as described by the INT data quality control system, and the
nightly photometric zeropoints provided by the ING were applied.

\item[$\bullet$] UKIRT  Telescope:\\
The cluster was observed in the near-infrared (near-IR) with the {\it Wide Field CAMera \citep[WFCAM,][]{2007A&A...467..777C}} in the course of the UKIRT InfraRed Deep Sky Surveys \citep[UKIDSS,][]{UKIDSS}. The UKIDSS survey provides a homogeneous coverage of the association in the Z,Y,J,H and Ks filters. The UKIDSS release (DR9) includes observations performed between September 2005 and January 2011, and are described in \citet{2007MNRAS.380..712L} and \citet{2012MNRAS.422.1495L}. We noticed that the Point Spread Function (PSF) of the pipeline processed interleaved images was not optimal for an accurate astrometric analysis. We therefore retrieved the individual frames from the WFCAM Science Archive \citep{WSA}. These frames are flat-fielded, dark subtracted, and sky-subtracted, and include an approximate astrometric solution with an accuracy better than a few arcsec. After extracting the sources from these individual images, a photometric calibration was derived using the UKIDSS catalogue.

\item[$\bullet$] KPNO Mayall Telescope:\\
We searched and retrieved  {\it NOAO Extremely Wide Field Infrared Imager} (NEWFIRM) \citep[][]{2003SPIE.4841..525A} and  {\it MOSAIC-1} \citep[][]{2000SPIE.3965...80W}  images in the NOAO Science Archive. The MOSAIC1 images were processed following standard procedures using the \emph{mscred} package within IRAF\footnote{IRAF is distributed by the National Optical Astronomy Observatory, which is operated by the Association of Universities for Research in Astronomy (AURA) under cooperative agreement with the National Science Foundation.} and the relevant calibration frames, as recommended in the User's manual. The detrended and sky-subtracted NEWFIRM images and their respective confidence maps were retrieved from the NOAO archive \citep{2009ASPC..411..506S}. The NEWFIRM $J$-band photometry was then tied to the UKIDSS one.

\item[$\bullet$] CTIO Blanco Telescope:\\
{\it MOSAIC2} is a clone instrument of the {\it MOSAIC1} installed on the Blanco telescope at CTIO. We retrieved the raw images from the NOAO Science Archive, and processed them following standard procedures using the \emph{mscred} package within IRAF and the relevant calibration frames, as recommended in the User's manual. 

\end{itemize}

Although in most cases sets of several consecutive and dithered images were
obtained, we chose to perform the analysis on the individual images rather than
on stacks. While individual images do not go as deep as stacks, this choice
ensures that the PSF (and hence the astrometric accuracy) and noise are not
affected by the stacking process. As we will see, using individual images offers
other advantages, in particular a more efficient rejection of problematic frames
or measurements, and an opportunity to reach out for faster moving objects.

\section{New observations \label{newobs}}
To complement the archival data and increase both the time baseline and spatial
coverage, we obtained deep wide field images of the cluster with {\it MegaCam}
at the {\it CFHT}. The observations were designed to optimize the astrometric
calibration. The various pointings were chosen to overlap by a few arcminutes,
ensuring an accurate astrometric anchoring over the entire survey. Each pointing
was obtained in dither mode, with a dither width of a few arcminutes. This
dithering allows filling the CCD-to-CCD gaps and correcting for deviant pixels
and cosmic ray events, and helps deriving an accurate astrometric solution over
the entire field-of-view (see section~\ref{chap:astsol}). The nights were
photometric, with an average seeing Full-Width at Half Maximum (FWHM) in the
range 0\farcs5--0\farcs7 as measured in the images. The data were processed and
calibrated with the recommended {\it Elixir} system, and the nightly magnitude
zero-points measured by the CFHT team were applied. 

\section{Observation properties}
\begin{figure}
   \centering
   \includegraphics[width=0.45\textwidth]{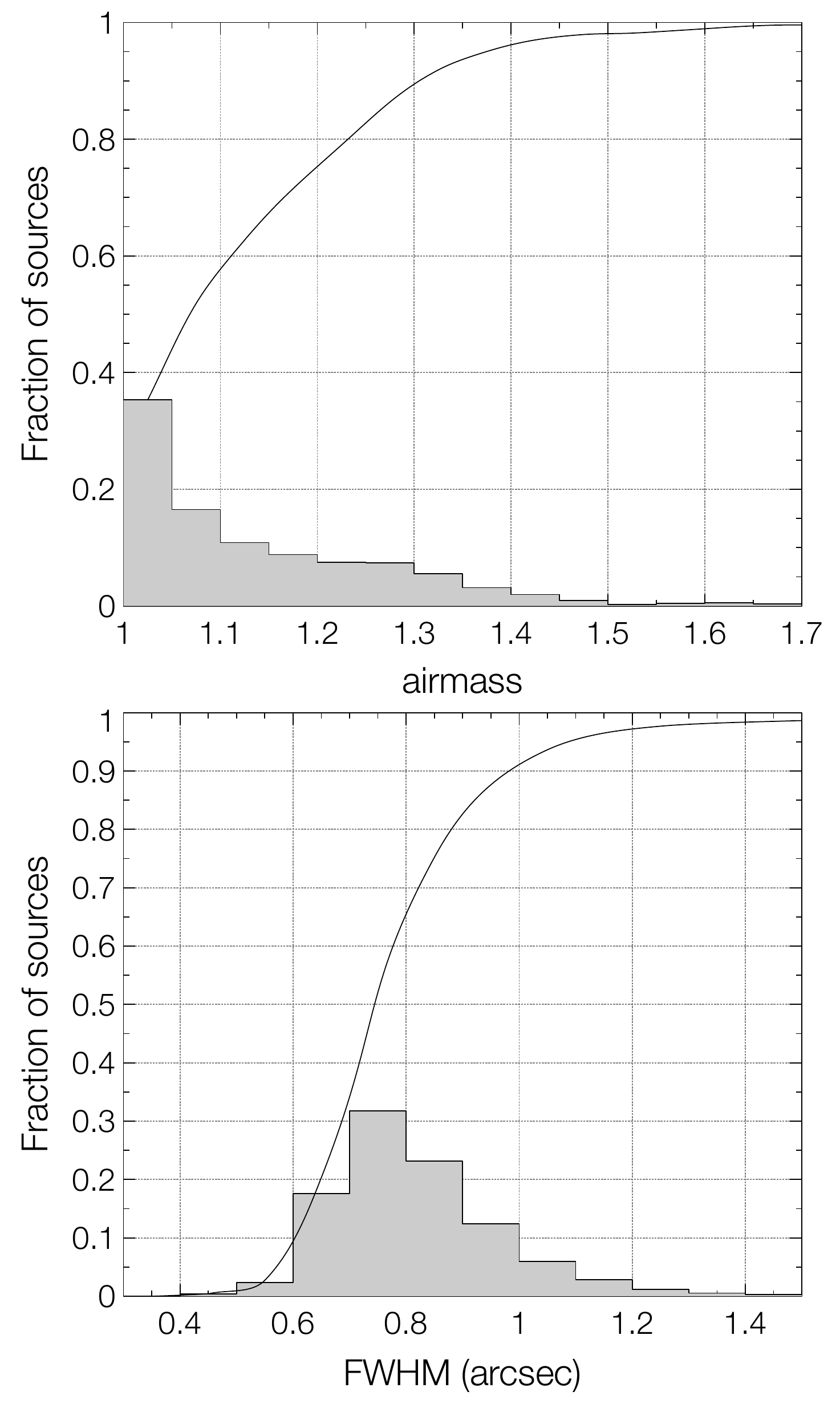}
      \caption{Upper panel: distribution of airmass for the observations used in this study. Lower panel: distribution of FWHM of all the individual unresolved detections. The lines represents the cumulative distribution.}
         \label{airmass_fwhm}
   \end{figure}

Describing the properties of every single individual image used in this study
woud be impractical. Instead, we present general statistics of 3 observational
properties especially important for the purpose of our study: airmass,
image FWHM and sensitivity. The latter two are important parameters as the best
positional accuracy achievable is mostly limited by the signal-to-noise ratio
(hence sensitivity), the FWHM of the point sources and sampling (pixel scale)
of the images \citep{1983PASP...95..163K}. The airmass is playing an important
role as well, as atmospheric turbulence and differential chromatic refraction
quickly increase with airmass. Figure~\ref{airmass_fwhm} shows the distribution
of airmass for the observations used in this study. About 75\% of the
observations were obtained at airmass $<$1.2, and $\approx$90\%  at airmass
$<$1.3. It also shows the distribution of the FWHM measured for all individual
unresolved detections (point sources). About two thirds have FWHM$\le$0\farcs8,
and about 92\%  have FWHM$\le$1\arcsec. Finally, even though the sensitivity of the individual frames varies greatly, the various observations
routinely reached luminosities fainter than 22~mag in the optical
($\lambda<$1.0~$\mu$m), and 18~mag in the near-IR ($\lambda>$1.0~$\mu$m). 
 
Whenever we could assess that an observation had been performed under good
photometric conditions and that an absolute photometric calibration (photometric
standard field) was available, we applied the corresponding zero-point to the
photometry extracted by {\sc SExtractor}. The associated absolute photometric
uncertainties are typically of the order of 5$\sim$10\%. Most
of the photometric measurements (all except the Subaru/SuprimeCam,
CTIO/MOSAIC2, KPNO/MOSAIC1 and some INT/WFC) were obtained under clear or
photometric conditions.

\begin{table*}
%\centering
\caption{Instruments used in this study and references to the corresponding surveys\label{table_obs}}

\begin{tabular}{lccccccclc}\hline\hline
Observatory   & Instrument        & Filters              & Platescale     & Chip layout   & Chip size  & Field of view & Survey Epoch      & Survey Area & Ref. \\
                      &                          &                         & [pixel$^{-1}$] &                      &                  &                     &                           & [deg$^{2}$] &      \\
\hline
CFHT              & UH~8K             & R, I                    & 0\farcs205 & 4$\times$2         &2k$\times$4k  & 29\arcmin$\times$29\arcmin & 1996               & 2.5  &  1 \\
CFHT              & CFHT~12K        & i                       & 0\farcs201 & 6$\times$2         & 2k$\times$4k & 42\arcmin$\times$28\arcmin & 1999               & 2.0  & 2,3 \\
CFHT              & MegaCam         & u,g,r,i               & 0\farcs187 & 4$\times$9         & 2k$\times$4k & 1\degr$\times$1\degr & 2004--2011 & 30  & 4 \\
Subaru            & Suprime-Cam  & r,i                     & 0\farcs200 & 5$\times$2         & 2k$\times$4k &  34\arcmin$\times$27\arcmin & 2002,2007     & 8     &  4 \\
INT                 & WFC                 & U,g,r,i,z             & 0\farcs333 & 3$\times$1+1    & 2k$\times$4k &  34\arcmin$\times$34\arcmin & 1998--2006 & 29  &   4,5,6,7 \\ 
UKIRT             & WFCAM            & Y,Z,J,H,K            & 0\farcs400 & 2$\times$2         & 2k$\times$2k &  45\arcmin$\times$45\arcmin\tablenotemark{a} & 2005--2011 & 79  & 8 \\
KPNO (Mayall) & NEWFIRM         & J                        & 0\farcs400 & 2$\times$2         & 2k$\times$2k &  28\arcmin$\times$28\arcmin & 2009               & 10   & 4 \\
KPNO (Mayall) & MOSAIC1         & VR-broad          & 0\farcs51\tablenotemark{b}  & 4$\times$2 & 2k$\times$4k & 36\arcmin$\times$36\arcmin & 2001--2003 & 8.5 & 4 \\
CTIO (Blanco)  & MOSAIC2         & VR-broad          & 0\farcs53\tablenotemark{b}  & 4$\times$2 & 2k$\times$4k & 36\arcmin$\times$36\arcmin & 2005--2006 & 1.0 & 4 \\
\hline
\end{tabular}

References: (1) \citet{1998A&A...336..490B}, (2) \citet{2001A&A...367..211M}, (3) \citet{2003A&A...400..891M}, (4) This study, (5) \citet{2002MNRAS.335..687D}, (6) \citet{2001NewAR..45...97M}, (7) \citet{1999A&AS..134..537Z}, (8) \citet{UKIDSS}

\tablenotetext{a}{with gaps of 12\farcm8 between each chip}
\tablenotetext{b}{images obtained using 2$\times$2 binning. Native pixel scale is half.}
\end{table*}

\section{Astrometric analysis \label{analysis}}

The astrometric analysis involves vast amounts of multi-epoch, multi-instrument,
multi-wavelength datasets, and requires highly automatized tools; all our
processing is done using the
Astr{\it O}matic\footnote{\url{http://www.astromatic.net}} software suite
\citep{2010jena.confE.227B}. The whole process is decomposed into the following
steps, which we describe in detail in the next sections:
\begin{enumerate}
\item Recovering and equalizing image metadata
\item Modeling the PSF
\item Cataloging
\item Quality assurance
\item Estimating astrometric uncertainties
\item Computing a global astrometric solution
\item Robust fitting of individual source  proper motions
\end{enumerate}

\subsection{Recovering and equalizing image metadata}
The PSF modeling, source extraction and astrometric calibration tasks
rely on a handful of parameters which must be set before processing the data.
These parameters comprise detector gains, saturation levels,
the approximate position and scale of the pixel grids on the sky, dates and
times of observation, durations of exposure, airmass, filter wheel position and
instrumental setups that define the instrumental context for the astrometric
solution (see \S\ref{chap:astsol}).

Recovering and uniformizing these parameters proves to be a laborious undertaking;
one faces here not one but nine different mosaic instruments over many years
of operation. In practice, and despite decades of effort from the community to
promote the standardization of metadata description in FITS headers, each
instrument uses slightly different conventions, which often evolves during the
life of the instrument. Among all parameters, the detector saturation level
(required for excluding saturated sources from the PSF modeling process and
from the astrometric solution) was found the less reliable. It was often
overestimated, and in one occasion it would ignore a scaling factor applied to
the data, a problem that also plagued the gains. Because of this we ended up
using SNR-vs-{\tt SPREAD\_MODEL} diagrams (see \S\ref{chap:qualcont} to
correct individual detector saturation levels.

\subsection{Modeling the Point Spread Function with {\sc PSFEx}}
The first step in making precise measurements of the positions of individual
sources is to compute an accurate model of the (variable) PSF for every
chip of every exposure.
A large fraction of the images (those with good seeing) are significantly
undersampled with some of the instruments, especially {\it WFCam} and {\it Mosaic2}. This
requires the PSF to be modeled at the sub-pixel level. The {\sc PSFEx} software
\citep{PSFEx} has been specifically designed to work with undersampled images
and arbitrary PSF shapes. Briefly, {\sc PSFEx} fits the image of every
point-source $\vec{p}_s$ with a projection on the local pixel grid of the linear
combination of basis vectors $\vec{\phi}_b$ by minimizing the $\chi^2$ function
of the coefficient vector $\vec{c}$
\begin{equation}
\label{eq:chi2psf}
\chi^2_{\rm PSF}(\vec{c}) = \sum_s \left(\vec{p}_s
		- \hat{\vec{p}}_s(\vec{c})\right)^T
		\mathbf{W}_s \left(\vec{p}_s - \hat{\vec{p}}_s(\vec{c})\right),
\end{equation}
where $\hat{\vec{p}}_s$ is the PSF model sampled at the location of star $s$:
\begin{equation}
\label{eq:psfmodel}
\hat{\vec{p}}_s(\vec{c}) =  f_s\,\mathbf{R}(\vec{x}_s)\sum_b\, c_b \vec{\phi}_b.
\end{equation}
$f_s$ is the flux within some reference aperture, and $\mathbf{W}_s$ the
inverse of the pixel noise covariance matrix for point-source $s$. We assume
that $\mathbf{W}_s$ is diagonal.
$\mathbf{R}(\vec{x}_s)$ is a resampling operator that depends on the image
grid coordinates $\vec{x}_s$ of the point-source centroid:
\begin{equation}
\mathbf{R}_{ij}(\vec{x}_s) =  h\left(\vec{x}_j
- \eta.(\vec{x}_i - \vec{x}_s)\right),
\end{equation}
where $h$ is a 2-dimensional interpolant (interpolating function), $\vec{x}_i$
is the coordinate vector of image pixel $i$, $\vec{x}_j$ the coordinate
vector of model sample $j$, and $\eta$ is the image-to-model sampling step
ratio (oversampling factor). We adopt a Lancz\'os-4 function \citep{duchon1979}
as interpolant. {\sc PSFEx} is able to model smooth PSF variations within each
chip by expanding the set of unknowns in (Eq. \ref{eq:psfmodel}) as a linear
combination of polynomial functions of the source position $\vec{x}_s$ in the
chip:
\begin{equation}
c_b = \sum_{k+l \le D} c_{b,k,l}\, x_1^{\,k}\,x_2^{\,l},
\end{equation}
where $D$ is the degree of the polynomial. We adopt $D=3$ (per chip), which is
found sufficient in practice to map PSF variations with the desired level of
accuracy for all the chips of all instruments involved here.

\begin{figure*}
   \centerline{
   \includegraphics[height=4.7cm]{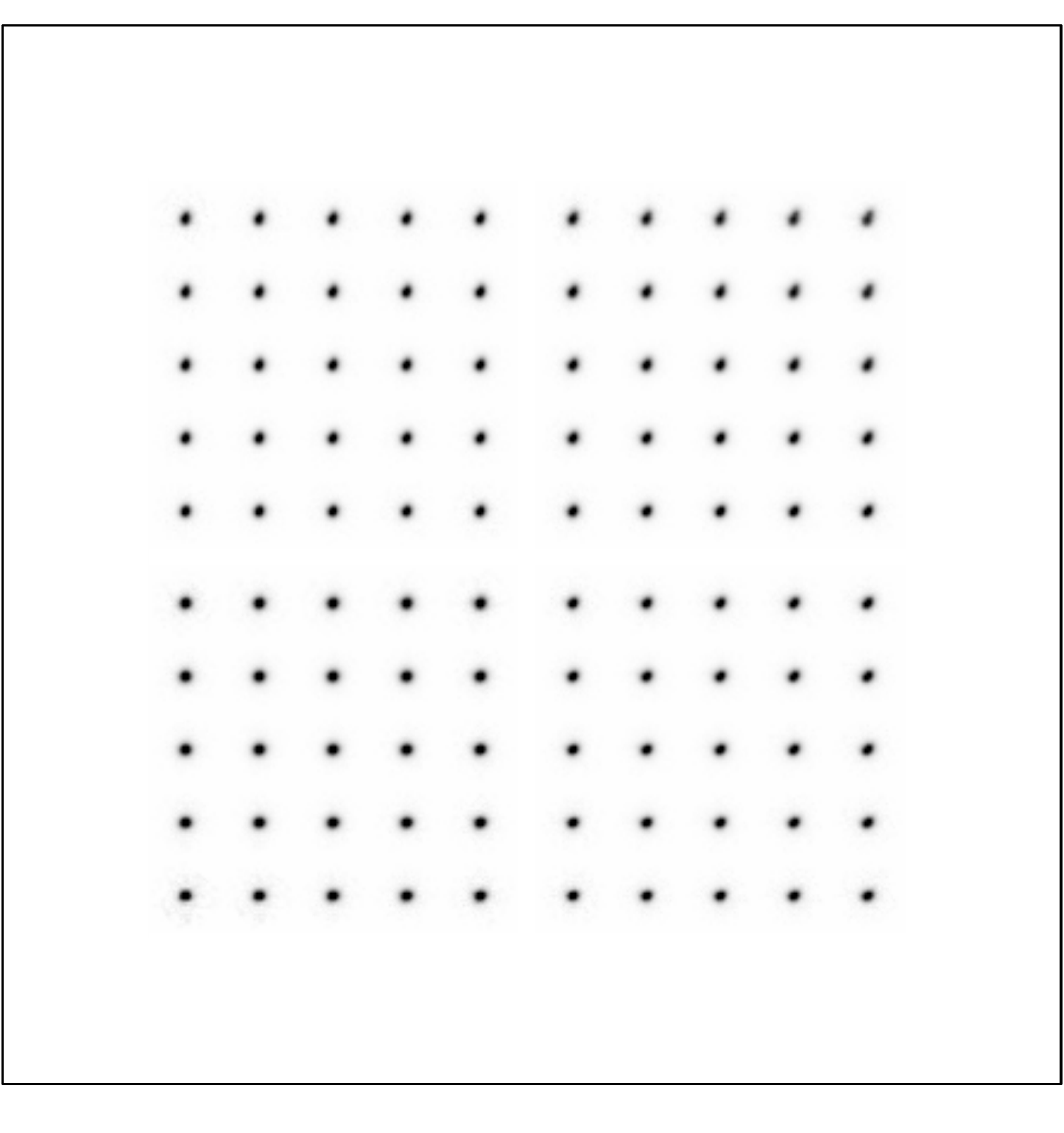}
\hspace{0.02\textwidth}
   \includegraphics[height=4.7cm]{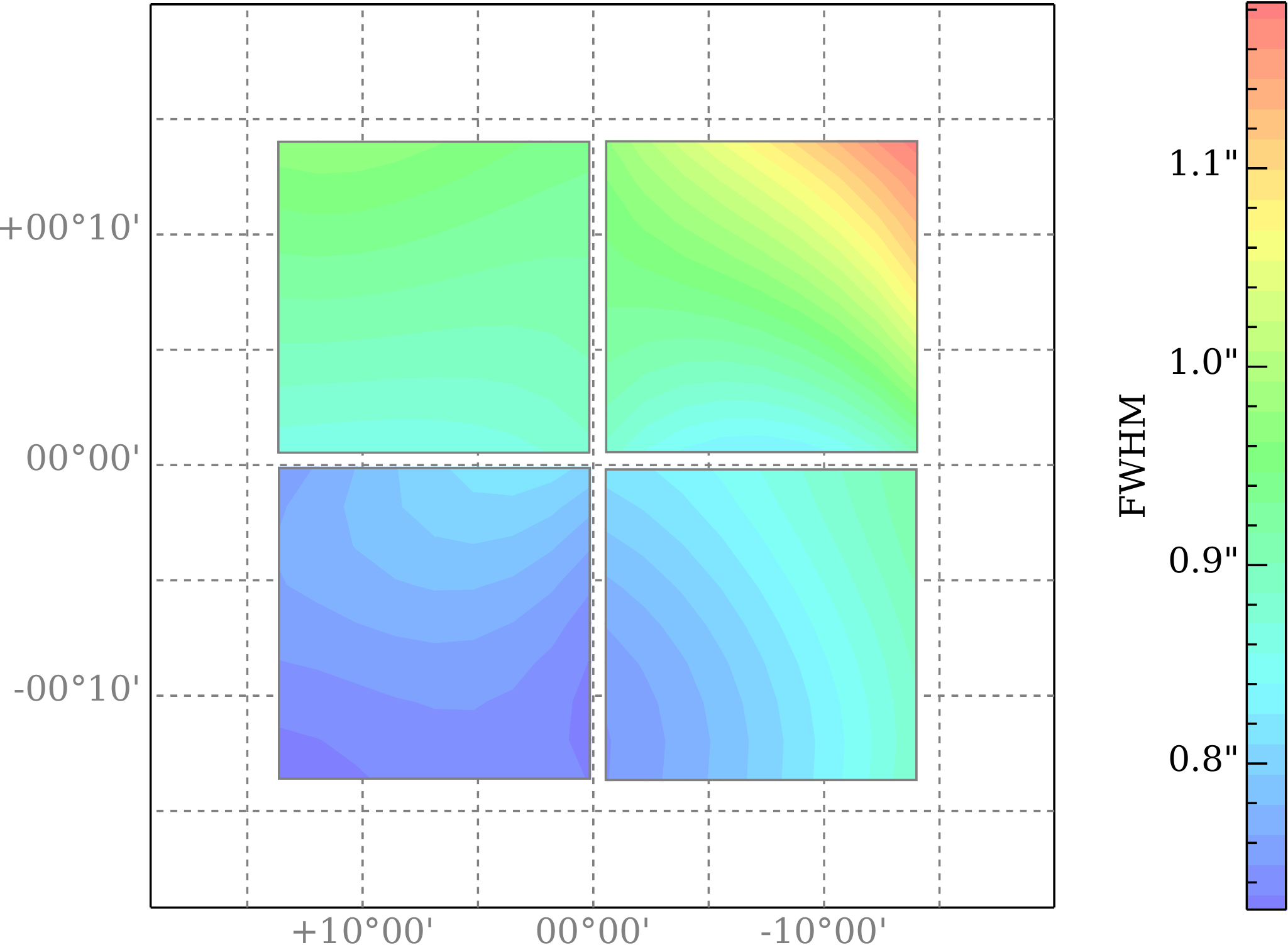}
\hspace{0.02\textwidth}
   \includegraphics[height=4.7cm]{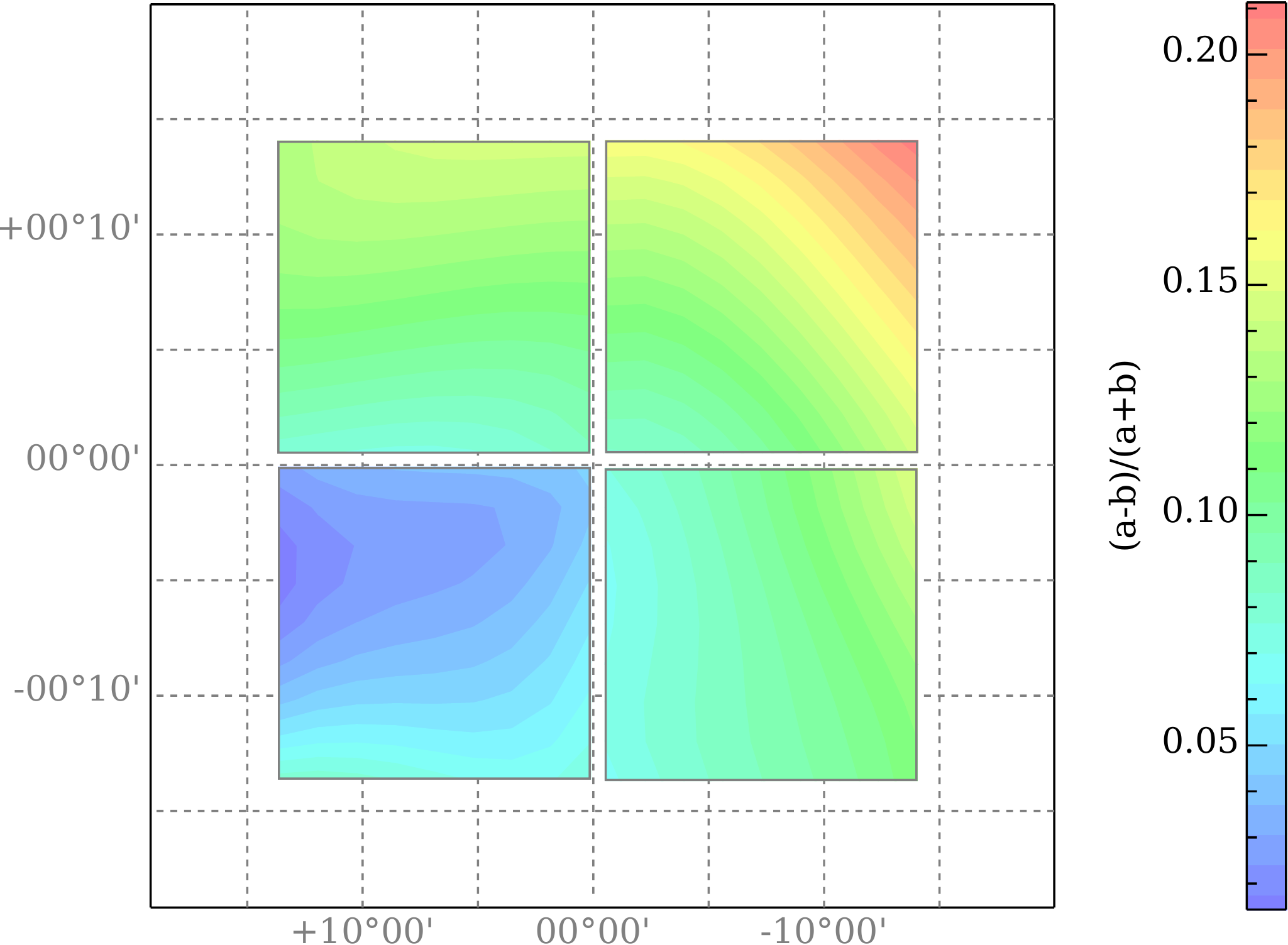}}
      \caption{Example of PSF reconstruction on one of the exposures
	(NEWFIRM frame {\tt K4N09B\_20091221-0011K-kp}).
	{\it Left:} snapshots of the variable PSF model over the field of view.
	{\it center:} distribution of the PSF model FWHM.
	{\it right:} distribution of the PSF model ellipticity.
	}
         \label{fig:psfexample}
   \end{figure*}

For this work we use the pixel basis as an image vector basis
$\vec{\phi}_b = \vec{\delta}(\vec{x}-\vec{x}_b)$, and therefore the $c_b$'s
directly represent pixel values of (super-resolved) images of the PSF. An
example of PSF model computed with {\sc PSFEx} is shown
Fig.~\ref{fig:psfexample}.

\subsection{Cataloging}
All sources with more than 3 pixels above 1.5 standard deviations of the local
background are extracted with the {\sc SExtractor} package
\citep{1996A&AS..117..393B}.
We measure fluxes and positions using the new S\'ersic model-fitting option
in {\sc SExtractor} \citep{PSFEx}, which relies on the empirical PSF model
previously derived by {\sc PSFEx}. In practice, PSF-convolved S\'ersic model
fits offer a level of astrometric accuracy comparable to that of pure PSF fits
for point sources, while making it possible to measure galaxy positions
(see \S\ref{chap:anchor}) and offering a better match to short asteroid
trails (see \S\ref{chap:asteroids}).
Contrary to fast iterative Gaussian centroiding (the so-called {\tt WIN}
estimates), they are largely immune to the spatial discretization effects
caused by undersampling. Moreover, model-fitting allows saturated pixels to be
censored without degrading excessively the positional accuracy on (moderately)
saturated stars, thereby significantly increasing the fraction of bright sources
suitable for astrometry.
Note that no extra-deblending of close pairs was attempted: a single,
PSF-convolved model was fitted to each detection.

\subsection{Quality assurance}
\label{chap:qualcont}
Not all archived exposures that match a given pointing location and the
desired range of seeing and airmass are acceptable for this study. Problems
such as tracking errors, bursts of electronic glitches, partially defocused
optical reflections (``ghosts'') and residual fringing patterns can alter
source centroids to a level that would affect significantly the computed proper
motions. All pre-selected exposures were therefore screened for defects using
semi-automated quality control based on {\sc PSFEx} and {\sc SExtractor}
measurements. By ``semi-automated quality control'' we mean automatically
generated statistics and plots prepared for human review
\citep[e.g.,][]{2004AN....325..583I,2006ASPC..351..731M,2006AJ....131.1163S,
2012ExA...tmp...19M}.

\begin{figure}
   \centering
   \includegraphics[width=\columnwidth]{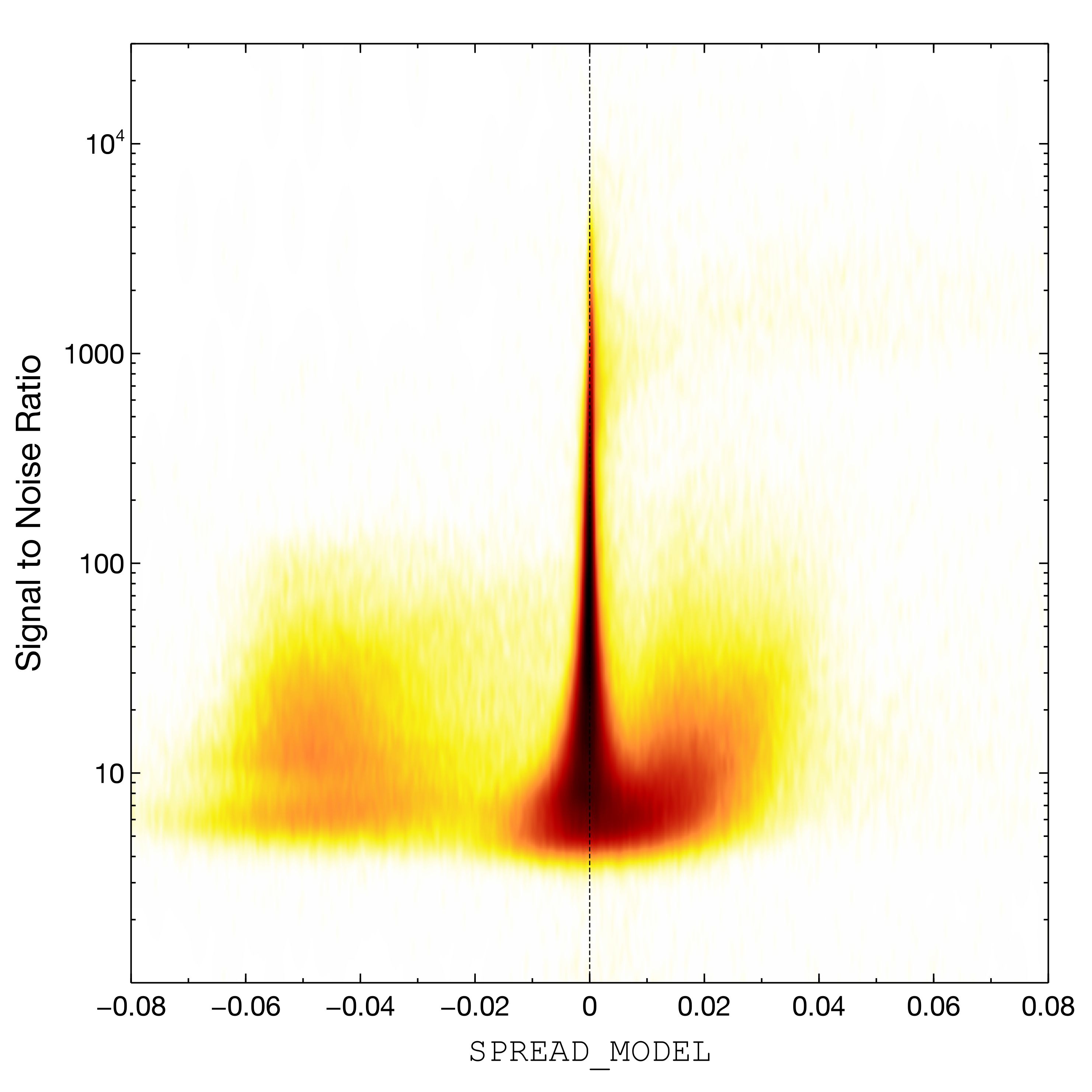}
   \caption{Density plot of signal-to-noise ratio (SNR) vs {\sc SExtractor}'s
	{\tt SPREAD\_MODEL} for all detections in this study. The
	dense vertical cloud located around {\tt SPREAD\_MODEL} is the point
	source (mostly stellar) locus. The fuzzy blob to the right of the
	stellar locus originates from galaxies and nebulosities, while the
	shallow cloud on the left is populated with cosmic ray hits and bad
	pixels. Note the asymmetry of the stellar locus caused by blended stars
	(most obvious at high SNRs).
	}
         \label{fig:spreadmodel}
   \end{figure}

Performing an extensive quality check in a large parameter space for 16,000
images coming from nine different mosaic instruments, each with its own
particular breed of issues, would be excessively time-consuming.
Instead we decided to focus on the consistency of the PSF, which all astrometric
measurements depend on. One way to check this consistency is to analyze
the distribution of the new {\tt SPREAD\_MODEL} estimator implemented in recent
development versions of {\sc SExtractor}, and originally developed as a
star/galaxy classifier for the Dark Energy Survey data management pipeline
\citep{2012arXiv1207.3189M,2012arXiv1204.1210D}. Briefly, {\tt SPREAD\_MODEL}
acts as a linear discriminant between the best fitting (local) PSF model
$\vec{\phi}$ derived with {\sc PSFEx} and a slightly ``fuzzier'' version made
from the same PSF model convolved with a circular exponential model with
scalelength = FWHM/16 (FWHM is the Full-Width at Half-Maximum of the local PSF
model). {\tt SPREAD\_MODEL} is normalized to allow comparing sources with
different PSFs throughout the field:
\begin{equation}
{\tt SPREAD\_MODEL} = \frac{\vec{\phi}^T {\bf W}\,\vec{x}}{\vec{\phi}^T {\bf W}
			\,\vec{\phi}}
	- \frac{\vec{G}^T {\bf W}\,\vec{x}}{\vec{G}^T {\bf W}\,\vec{G}},
\end{equation}
where $\vec{x}$ is the image centered on the source, and ${\bf W}$ the inverse
of its covariance matrix (which we assume to be diagonal). By construction,
{\tt SPREAD\_MODEL} is close to zero for point sources, positive for
extended sources (galaxies), and negative for detections smaller than the
PSF, such as cosmic ray hits. Figure \ref{fig:spreadmodel} shows the
typical distribution expected for source signal-to-noise ratio (SNR) as a
function of {\tt SPREAD\_MODEL}. We found this diagram to be extremely
effective at revealing a wide range of cosmetic and morphometric issues that
can arise with survey images:
\begin{itemize}
\item any significant departure of the point-source locus from a narrow
distribution centered on ${\tt SPREAD\_MODEL} = 0$ is a sign that the PSF model
does not fit properly point-sources. The reason may be a problem with the
PSF modeling process (e.g., the model cannot follow the variations of the PSF
throughout the field), a non-linear behavior of the detector (e.g., saturated
stars), excessive source confusion (very poor seeing), or a multi-modal PSF
({\sc SExtractor} identifies as multiple source different parts of the PSF, in
cases of strong defocusing or guiding errors for instance)
\item a burst of bad pixels or electronic glitches shows up as a denser
cloud on the left part of the diagram.
\item optical ``ghosts'', diffraction spikes or faint satellite track are
broken up into pieces by {\sc SExtractor} and appear as spots on the right part
of the diagram.
\item background inhomogeneities, such as contamination by strong fringe
residuals or large textured halos, produce a large horizontal blur in the lower
part of the diagram.
\end{itemize}

We visually inspected SNR vs {\tt SPREAD\_MODEL} plots for all 16,515 exposures
pre-selected for this survey. We identified and rejected 427 exposures (2.6\%)
exhibiting at least one of the signatures listed above, or lying amidst a
sequence of ``bad'' frames, and for which we judged
that quality issues were severe enough to harm the accuracy of astrometric
measurements: 243 were plagued mostly by electronic artifacts (all UKIDSS), 97
by guiding errors, 40 by optical ghosts, 13 by very bad seeing, 13 by excessive
fringing, 8 by defocused images, and 7 by various background issues. In
addition, 6 short NEWFIRM exposures with poor cosmetics did not have enough
``clean'' stellar images to derive a proper PSF model.  
 
\subsection{Estimating astrometric uncertainties \label{chap:astrom_error}}
Position uncertainties play a prominent role in our astrometric pipeline. They
are the main ingredients of the relative weights given to individual detections
in the global astrometric solution. They are also used in the computation of
robust proper motions, as weights, and to identify outliers.

{\sc SExtractor}'s 1-$\sigma$ fitting-error ellipse parameters
{\tt ERRAMODEL\_IMAGE}, {\tt ERRBMODEL\_IMAGE} and {\tt ERRTHETAMODEL\_IMAGE}
are directly extracted from the covariance matrix computed in the
{\sc LevMar} Levenberg-Marquardt minimization engine \citep{lourakis04LM}. Based
on repeatability tests performed with a wide range of simulated (photon-noise
dominated) images, we checked that the estimated uncertainties matched the
observed standard deviation of position residuals to better than 10\% for
isolated sources.

However the dominant source of positional uncertainties for bright stars on
ground-based exposures, with a duration of a few minutes or less, is not photon
noise, but apparent relative motion caused by atmospheric turbulence.
This motion is highly correlated at small angles \citep{1916PAllO...3....1S};
its impact on the estimation of proper motions is small when working with
very small fields of view, or when positions are measured relative to close
neighbors. Neither is the case here, and the contribution from atmospheric
turbulence component must be taken into account.
In the regime probed by these observations (exposure time, field-of-view,
telescope diameter), theoretical considerations as well as experimental
studies \citep{1980A&A....89...41L,1981PrOpt..19..281R,1989AJ.....97..607H,
1992A&A...262..353S,1995PASP..107..399H} have established that the amplitude of
the relative random motion between two sources separated by angle $\theta$
(in arcmin) is well described by
\begin{equation}
\label{eq:stellmot}
\sigma_m\,(\theta,T) = \sigma_{0m}\ (\theta/10)^{1/3}\ T^{-1/2},
\end{equation}
where $T$ is the exposure time in seconds, and $\sigma_{0m}$ is the standard
deviation expected in unit time for a pair of stars separated by ten arcmin.

Correlated ``position noise'' as described by Eq.~(\ref{eq:stellmot}) translates
into non-diagonal terms in the measurement error matrix of detections from
individual exposures. But since the current version of our astrometry solver
ignores non-diagonal terms in the weighting matrix, we are left with considering
only the variance averaged over individual fields.
Part of this variance is ``absorbed'' in the deformable distortion model
\citep{1978moas.coll..339C,1980A&A....89...41L} such as the second-degree
polynomial we are using (\S\ref{chap:astsol}), but we assume this dampening
effect to be small considering the wide fields of view of all the instruments
involved here.
The average contribution (per source) to pairwise positional variance due to
relative motions in an exposure is half the integral of Eq.~(\ref{eq:stellmot})
over all possible pairs of positions within the field of view FOV:
\begin{equation}
\label{eq:intstellmot}
\sigma^2_M({\rm FOV},T)
	= \frac{1}{2} \int_{FOV} d\vec{\theta}_1 \int_{FOV} d\vec{\theta}_2\ 
	\sigma^2_m\,(||\vec{\theta}_1 - \vec{\theta}_2||,T).
\end{equation}
For rectangular FOVs, we find that the following expression provides a good
approximation (within 5\% for aspect ratios $< 20:1$) to $\sigma_M({\rm FOV})$:
\begin{equation}
\label{eq:apintstellmot}
\sigma_M({\rm FOV},T) \approx \frac{1}{\sqrt{2}}\ 
		\sigma_{0m} \left(\frac{\theta_{\,\rm FOV}}{30'}\right)^{1/3}
		\ T^{-1/2},
\end{equation}
where $\theta_{\,\rm FOV}$ is the diagonal of the field in arcmin. 

Using star trails, \cite{1995PASP..107..399H} measure $\sigma_{0m} = 54$ mas at
Mauna Kea, whereas \cite{1989AJ.....97..607H} report a much higher
$\sigma_{0m} = 143$ mas at Allegheny observatory in Pittsburgh.
\cite{1996PASP..108.1135Z}, analyzing astrometric calibration residuals from
short, repeated observations made at Kitt Peak and Cerro Tololo, found results
compatible in average with Han \& Gatewood's value; although he claims that
their exposures with best seeing exhibit twice less dispersion, and hint at
a dependency of turbulence-induced motion with seeing. Our own
measurements using short wide-field exposures from archive data (Bouy et al.,
in preparation), exhibit litte dependency on actual seeing, and suggest that
Han \& Gatewood's value is appropriate for observations carried out in good
sites.
We therefore add $\sigma_M$ in quadrature to the measurement uncertainties
estimated by {\sc SExtractor}, adopting $\sigma_{0m} = 54$ mas as well as the
FOV and exposure time of the current image. Note that the current version of
our astrometry engine assumes that position uncertainties are isotropic.

Another source of errors in the measurement of positions is imperfect
deblending of close detections. Of particular concern for the astrometric
solution are the detrimental effects of deblending errors in some bright
sources. The impact of deblending on centroid measurements varies a lot from
object to object and is difficult to quantify {\em a priori}.
Nevertheless we find that adding a $0.1$ pixel error in quadrature to position
uncertainties of detections flagged as ``deblended'' by {\sc SExtractor}
alleviates the issue with the bright sources, without downweighting
excessively sources that have been properly deblended.

\subsection{Computing a global astrometric solution}
\label{chap:astsol}
The global astrometric solution is computed with version 2.0 of the
{\sc SCAMP} software package \citep{2006ASPC..351..112B}. {\sc SCAMP} is itself
a mini-pipeline performing various operations before and after computing the
global solution {\it  per se}. These operations are described in details in the
{\sc SCAMP} documentation; in the following we focus only on those that are
especially important for this study.

The global solution computed by {\sc SCAMP} is the result of minimizing the
quadratic sum of differences in position between overlapping detections from
pairs of catalogs, an approach pioneered by \cite{1960AN....285..233E}:
\begin{equation}
\label{eq:chi2astrom}
\chi^2_{\rm astrom} = \sum_s\sum_a\sum_{a>b}\ 
	\frac{1}{\sigma^2_{s,a} +\sigma^2_{s,b}}\ 
	||\vec{\xi}_a(\vec{x}_{s,a}) - \vec{\xi}_b(\vec{x}_{s,b})||^2,
\end{equation}
where $s$ is the source index, $a$ and $b$ are catalog indices,
and $\sigma_{s,a}$ is the positional uncertainty for source $s$ in
catalog $a$. For the purpose of computing a global solution,
positions in Eq.~(\ref{eq:chi2astrom}) are in a common system of
{\it reprojected} coordinates derived from raw detector coordinates $\vec{x}$.
For mosaic cameras, a catalog comprises several sub-catalogs for each exposure:
one per detector chip. We express the reprojection operator $\vec{\xi}_{c,e}$
for chip $c$ and exposure $e$ as a combination of an undistorted reprojection
operator $\vec{\xi}^0_{c,e}$ derived from the (tangential) projection
approximated at the initial cross-matching stage, and two polynomials describing
instrumental distortions:
\begin{equation}
\label{eq:xi}
\vec{\xi}_{c,e}(\vec{x}) = \vec{\xi}^0_{c,e}\left(\vec{x}
	+ \sum_{p} \vec{f}_{c,i,p} \phi_p(\vec{x})
	+ \sum_{m} \vec{g}_{e,m} \psi_m(\vec{\rho})\right).
\end{equation}
The first polynomial with free coefficients $\vec{f}_{c,i,p}$ describes static,
chip-dependent ($c$) and instrument-dependent ($i$) distortions that are
function of raw coordinates $\vec{x}$. The second polynomial with free
coefficients $\vec{g}_{e,m}$ accounts for exposure-dependent distortions that
are function of focal-plane coordinates $\vec{\rho}$, computed from the
raw coordinates $\vec{x}$ using the initial positioning of chips on the focal
plane. In this study we adopt a degree 4 for the chip-dependent polynomial,
which in practice provides a very good fit to the geometrical distortions of
most instruments. We choose a degree 2 for the exposure-dependent polynomial to
account for flexures and geometric atmospheric refraction.
Note that {\sc SCAMP} automatically and progressively reduces
the degree of both polynomials in cases where the number of free parameters
reaches or exceeds the number of constraints: detector failures, shallow
exposures, etc.

The cameras involved in this study are often taken off the telescope between
runs. Experience shows that the static part of the distortion pattern
changes from run to run, sufficiently enough to undermine the global solution.
The same goes with filter changes. Therefore the count of ``astrometric
instruments'' entering Eq.~(\ref{eq:xi}) far exceeds that of cameras, because
what matters is eventually the combination camera/filter/run. Relying on
header information and logbooks, we identified 94 such combinations for
the whole dataset, taken with 9 cameras through 30 filters.

For the minimum of $\chi^2_{\rm astrom}$ to be unique, the solution must be
anchored on the sky. {\sc SCAMP} does that by forcing one of the catalogs in
Eq.~(\ref{eq:chi2astrom}) to be a catalog of astrometric references with
fixed \vec{\xi} coordinates. We selected 2MASS \citep{2006AJ....131.1163S} as a
reference catalog, because of its suitable depth, good homogeneity, and tight
range of observation epochs.

Because instrumental distortions are small at the scale of a chip (typically a
few pixels), Eq.~(\ref{eq:xi}) can be well approximated by
\begin{equation}
\label{eq:xiapp}
\vec{\xi}_{c,e}(\vec{x}) \approx \vec{\xi}^0_{c,e}(\vec{x}) +
	\jacob{\vec{\xi}^0}{\vec{x}}{\vec{x}}\ 
	\left(\sum_{p} \vec{f}_{i,p} \phi_p(\vec{x})
	+ \sum_{m} \vec{g}_{e,m} \psi_m(\vec{\rho})\right),
\end{equation}
which makes the minimization of Eq.~(\ref{eq:chi2astrom}) equivalent to solving a
system of linear equations.

   \begin{figure*}
   \centerline{
   \includegraphics[width=0.5\textwidth]{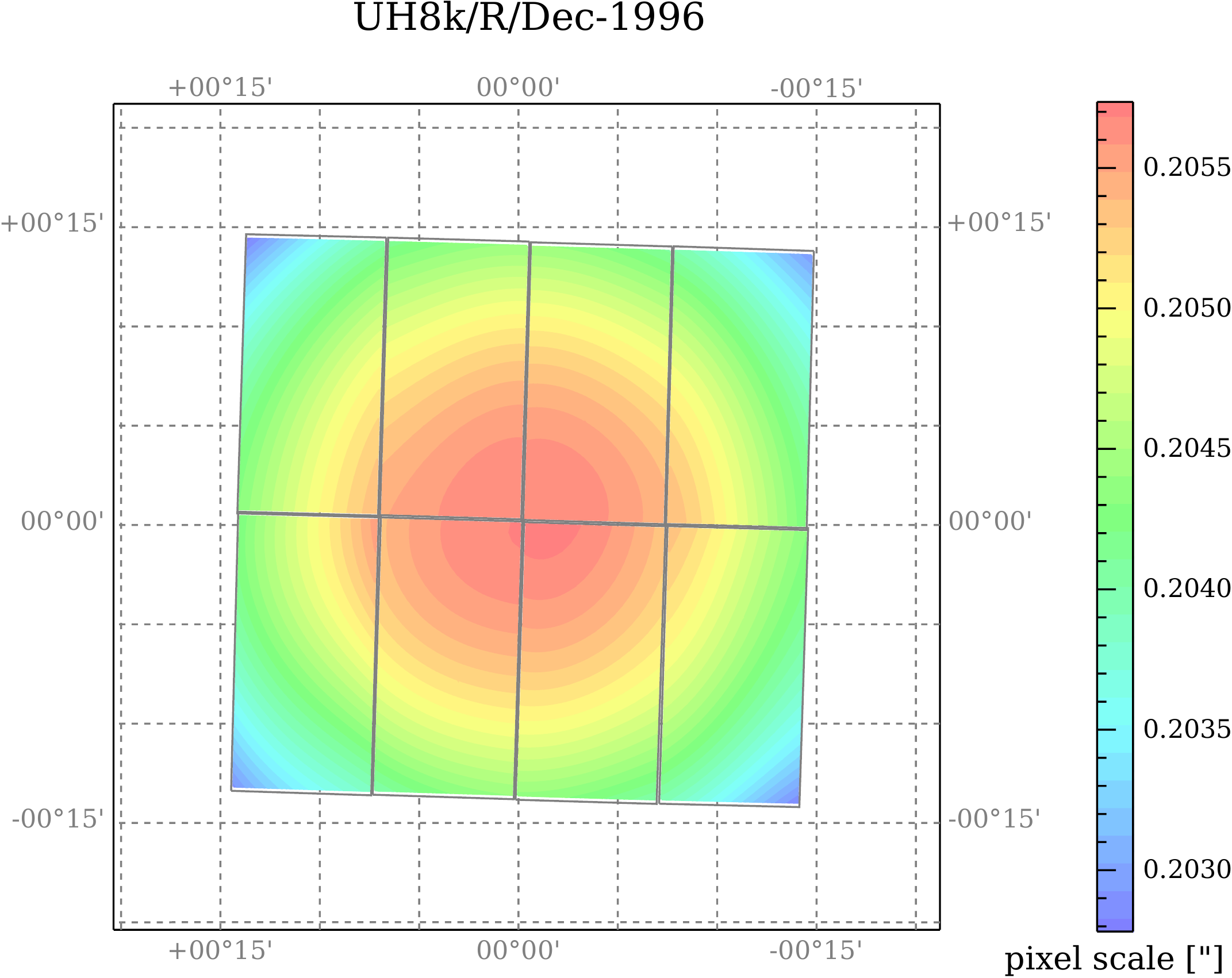}
   \includegraphics[width=0.5\textwidth]{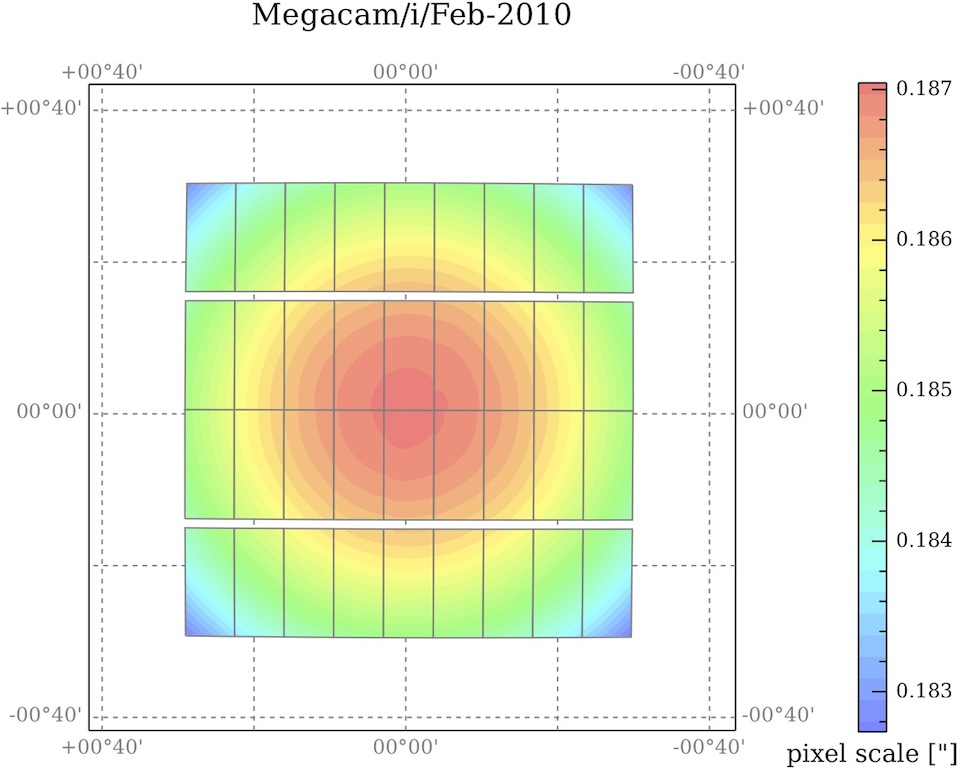}}
\vspace{5mm}
   \centerline{
   \includegraphics[width=0.5\textwidth]{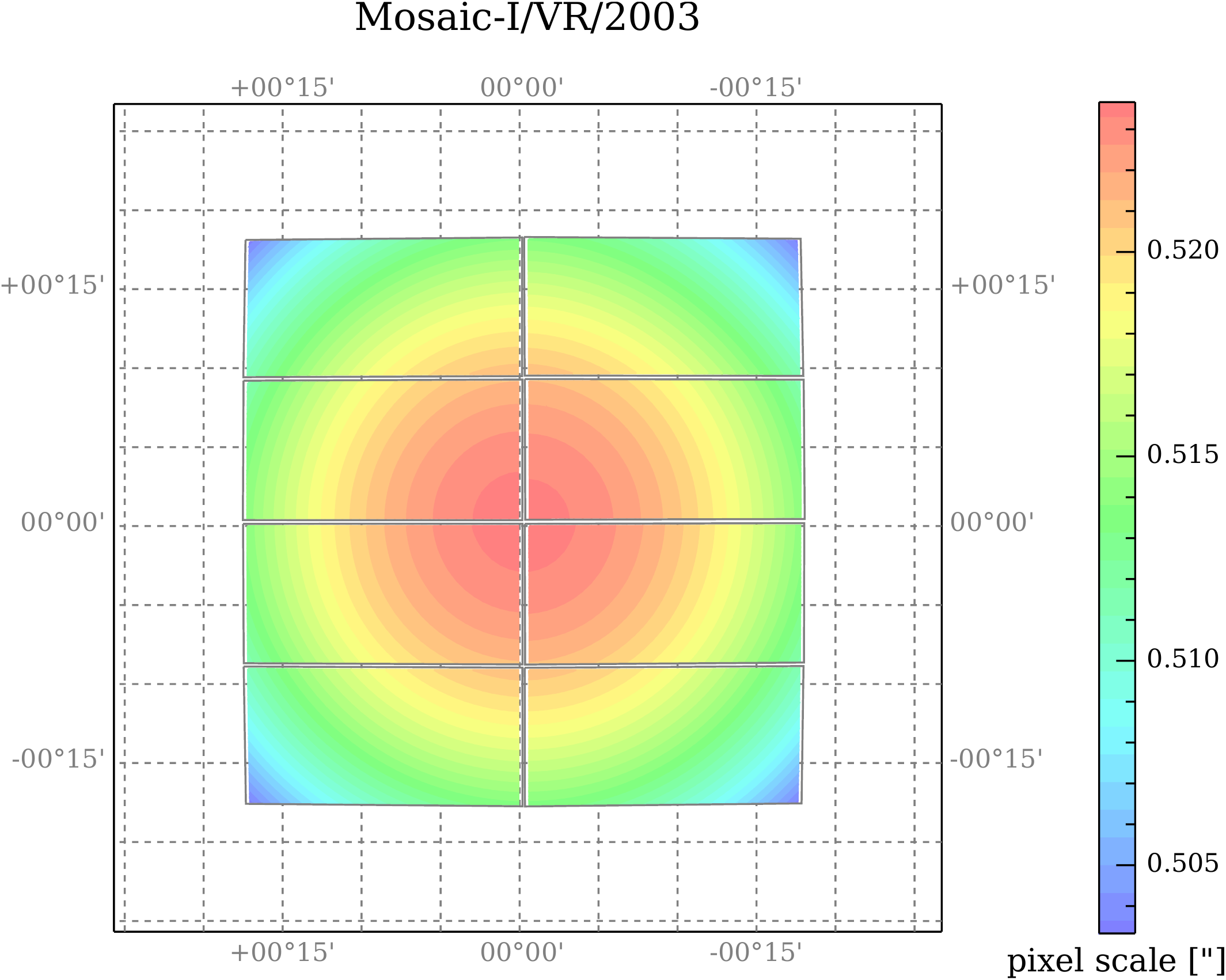}
   \includegraphics[width=0.5\textwidth]{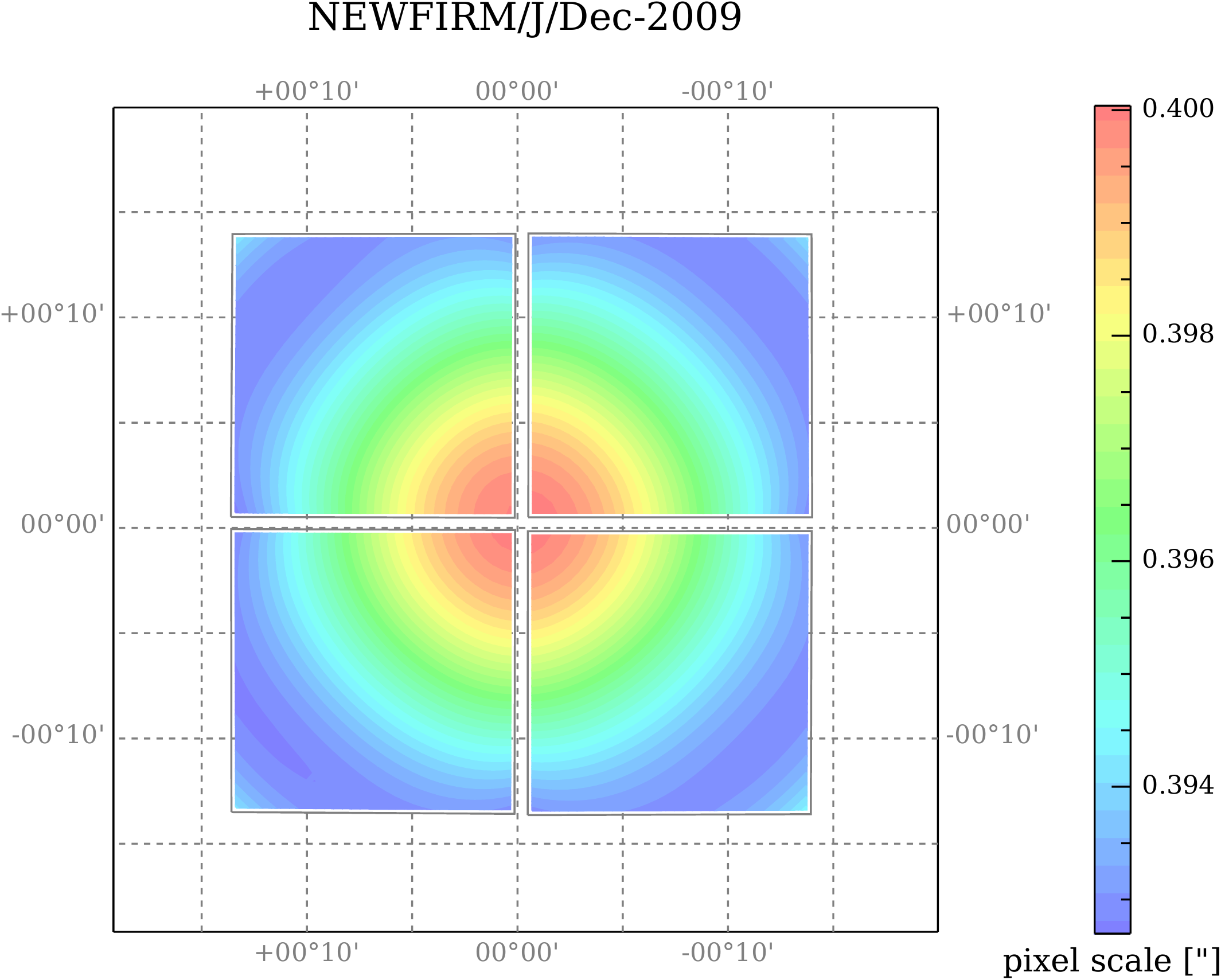}}
\vspace{5mm}
   \centerline{
   \includegraphics[width=0.5\textwidth]{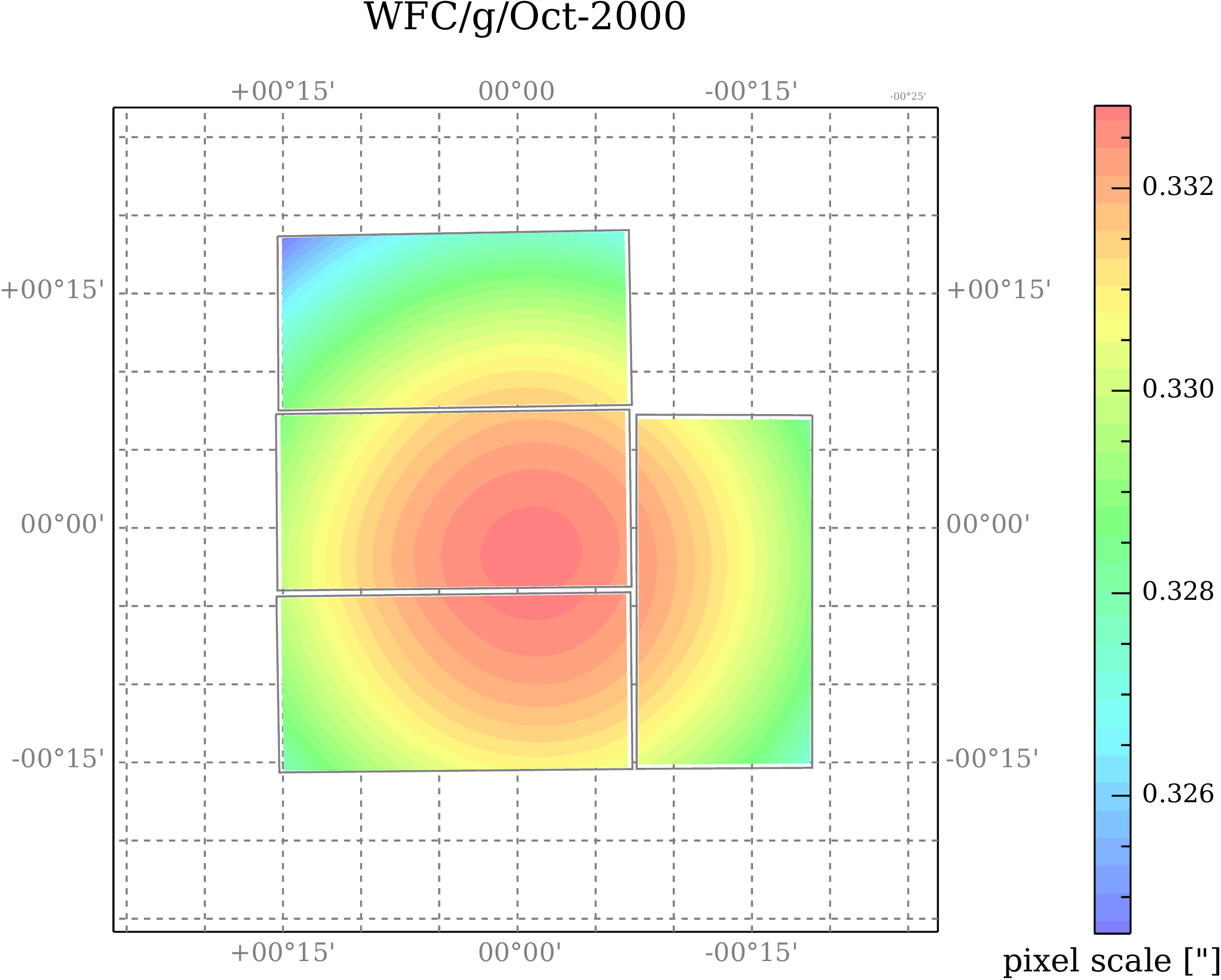}
   \includegraphics[width=0.5\textwidth]{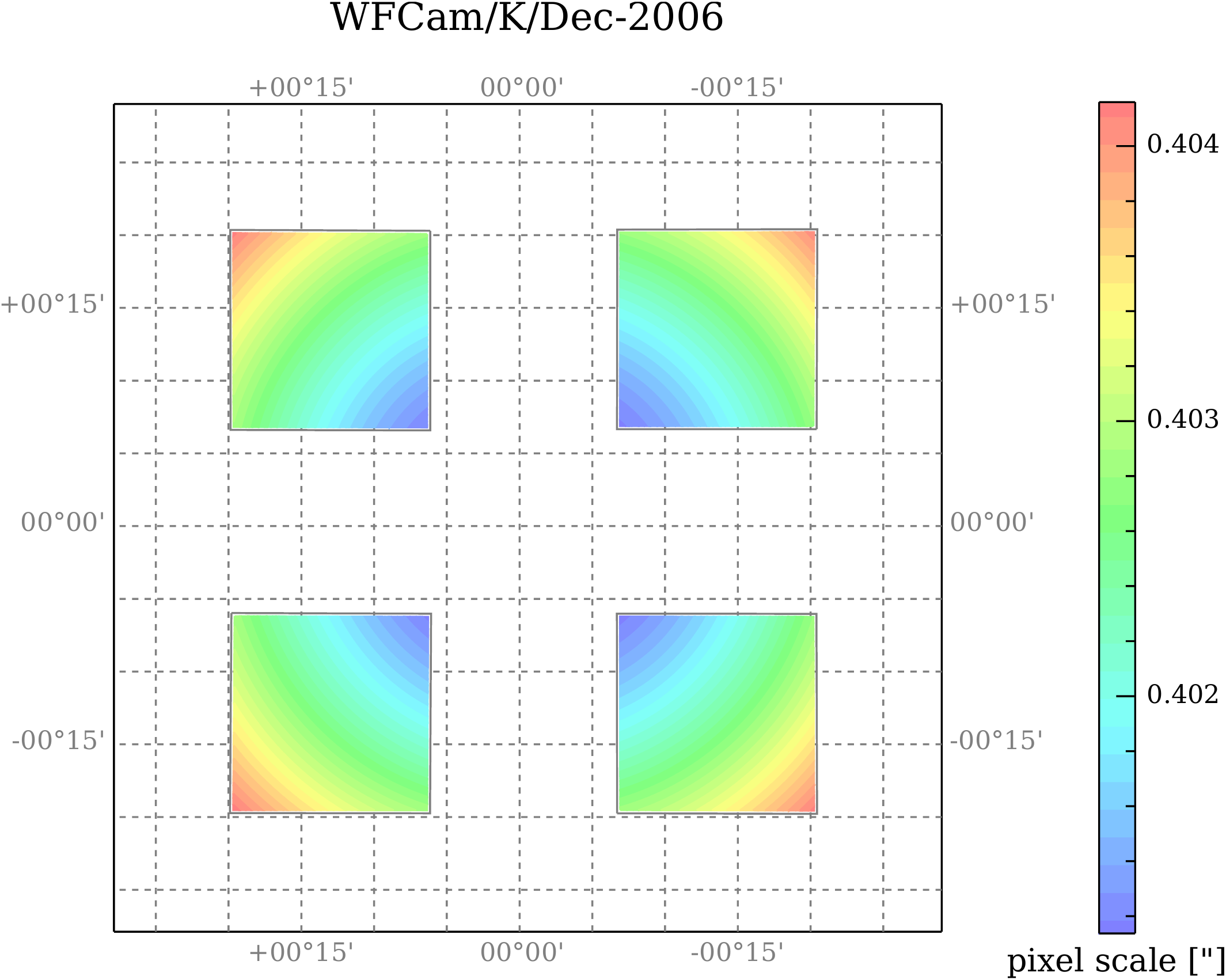}}
   \caption{Examples of camera distortion patterns, represented by maps of
	the pixel scale, for 6 of the 94 camera/filter/run combinations.
	}
   \label{fig:distort}
   \end{figure*}

In {\sc SCAMP}, the astrometric solution is computed three times. A first,
distortion-free solution accurate to about 1" is obtained from the registration
of all image exposures. This is sufficient to provide a satisfactory matching
of most overlapping detections. A full global solution is then computed, which
is used to identify detections with calibrated positions deviating excessively
from the mean. Strong deviations may be caused by cross-matching issues
such as blending or mismatches, differential chromatic refraction (at high
airmass), wavelength-dependent centroids (in galaxies), or large proper motions
(in stars). Because of the extended range of epochs and the presence of a nearby
star cluster in the data, we opt for a somewhat severe level of clipping,
rejecting about 4.5\% of all detections at this stage. The final run of the
solver on this clipped sample yields the final set of distortion parameters.
Figure \ref{fig:distort} shows examples of recovered distortion patterns for
some of the 94 camera/filter/run combinations.

\begin{figure*}
   \centerline{
   \includegraphics[height=0.45\textwidth]{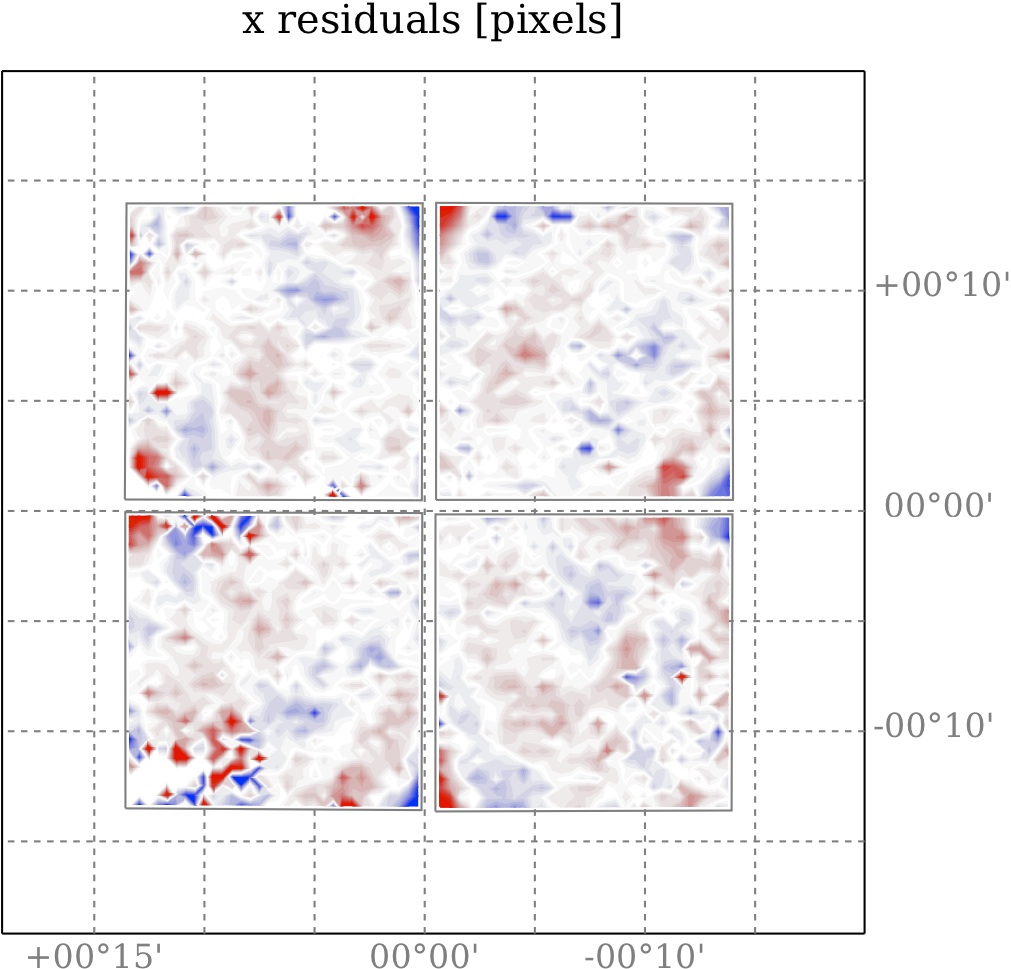}
   \includegraphics[height=0.45\textwidth]{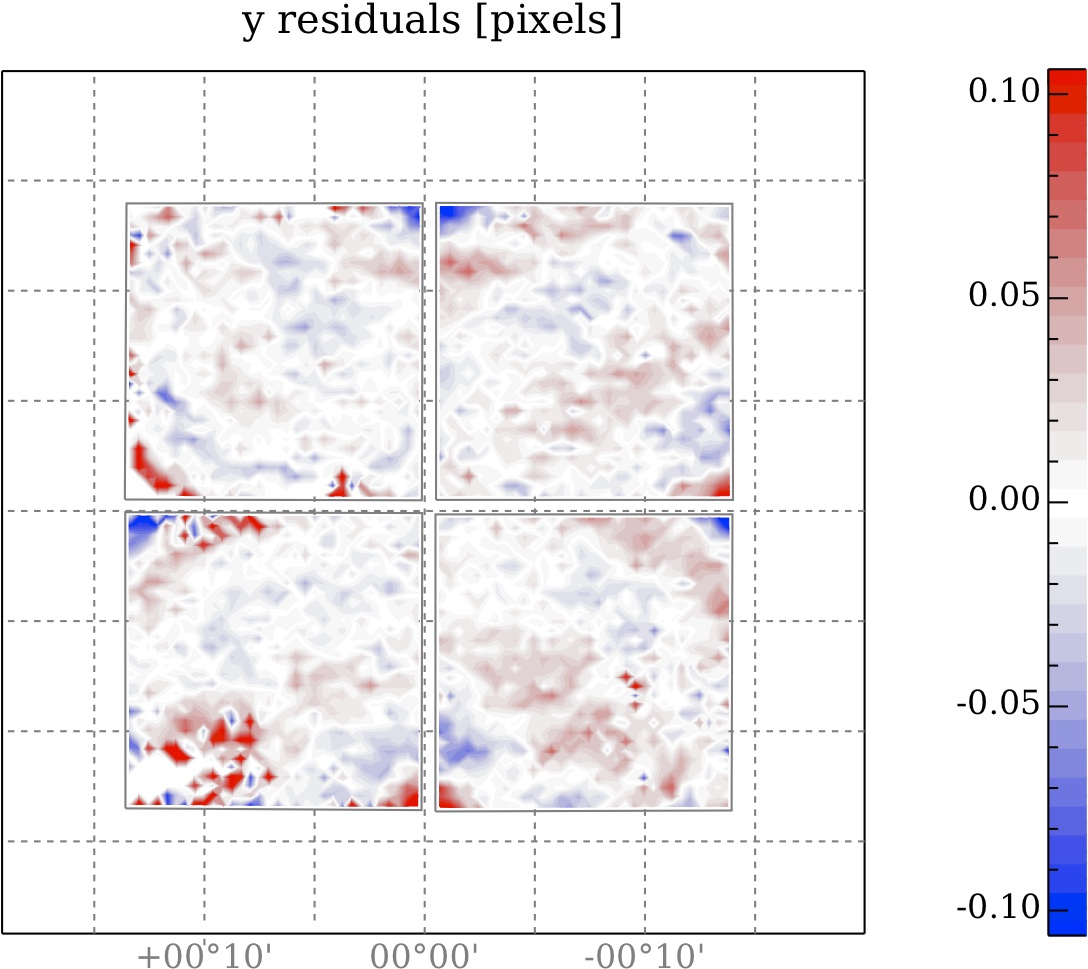}}
   \vspace{5mm}
   \centerline{
   \includegraphics[height=0.45\textwidth]{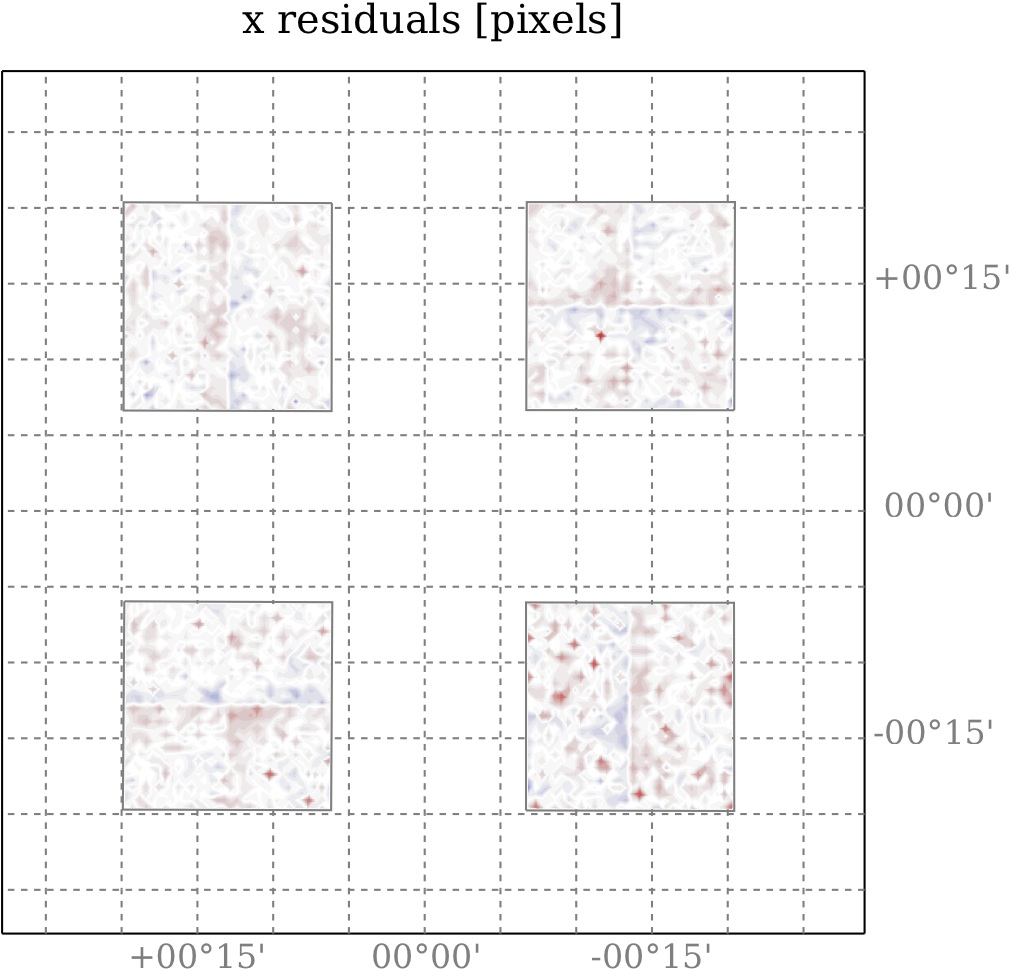}
   \includegraphics[height=0.45\textwidth]{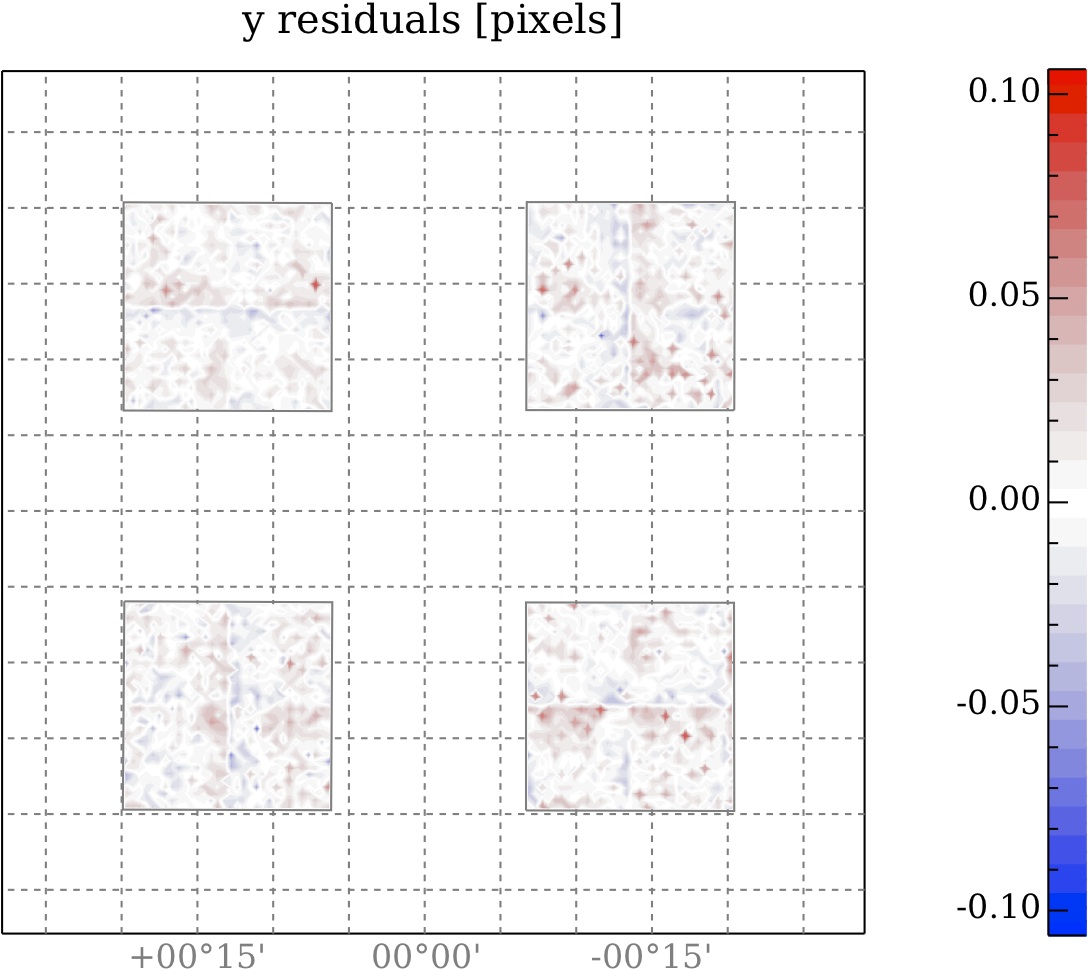}}
   \caption{$x$/$y$ maps of systematic residuals (in pixels) for two
	camera/filter/run combinations. {\it (Top):} NEWFIRM/J-band/Dec-2009.
	{\it (Bottom):} WFCam/Z-band/Dec-2005. }
   \label{fig:resi}
   \end{figure*}

{\sc SCAMP} offers the possibility to produce maps of the average residuals in
raw coordinates after calibrating the positions with the best-fitting
distortion pattern models. These maps tell us of possible position-dependent
systematic calibration errors, in particular distortion features that cannot
be fitted with a $4^{\rm th}$ degree polynomial. Two cameras appear to exhibit
particularly striking residual patterns with most filter/runs 
(Fig. \ref{fig:resi}). A periodic, symmetric pattern is seen for {\sc NEWFIRM}
with amplitude $\pm0.05$~pixel, an indication that a $4^{\rm th}$ degree
polynomial is a poor fit to the distortion profile of this instrument. The
{\sc WFCam} data show coordinate jumps up to $0.08$~pixel between the
$1024^{\rm th}$ and the $1025^{\rm th}$ rows and between the $1024^{\rm th}$ and
the $1025^{\rm th}$ columns. While the most obvious explanation to this feature 
would be small physical gaps between the four quadrants of the Hawaii-2
detectors \citep{2000SPIE.4028..331C}, this ``geometrical'' hypothesis was
dismissed by the Teledyne engineers we contacted, after a careful examination
of the original mask used to manufacture the arrays. At the time of writing
we remain clueless about the origin of this issue, which is virtually
undetectable using the UKIDSS data alone, because of the survey micro-dithering
and tiling strategy.

\subsection{Differential chromatic refraction \label{dcr}}
Dispersive elements along the optical path (atmosphere, lenses) have
wavelength-dependent refraction indices producing a color dependent shift of the
centroid \citep[e.g.,][]{1982PASP...94..715F, 1992AJ....103..638M}. This effect
is known as differential chromatic refraction (DCR). The magnitude of
atmospheric DCR depends on zenithal distance and on the source color index.

\begin{figure}
   \centering
   \includegraphics[width=0.45\textwidth]{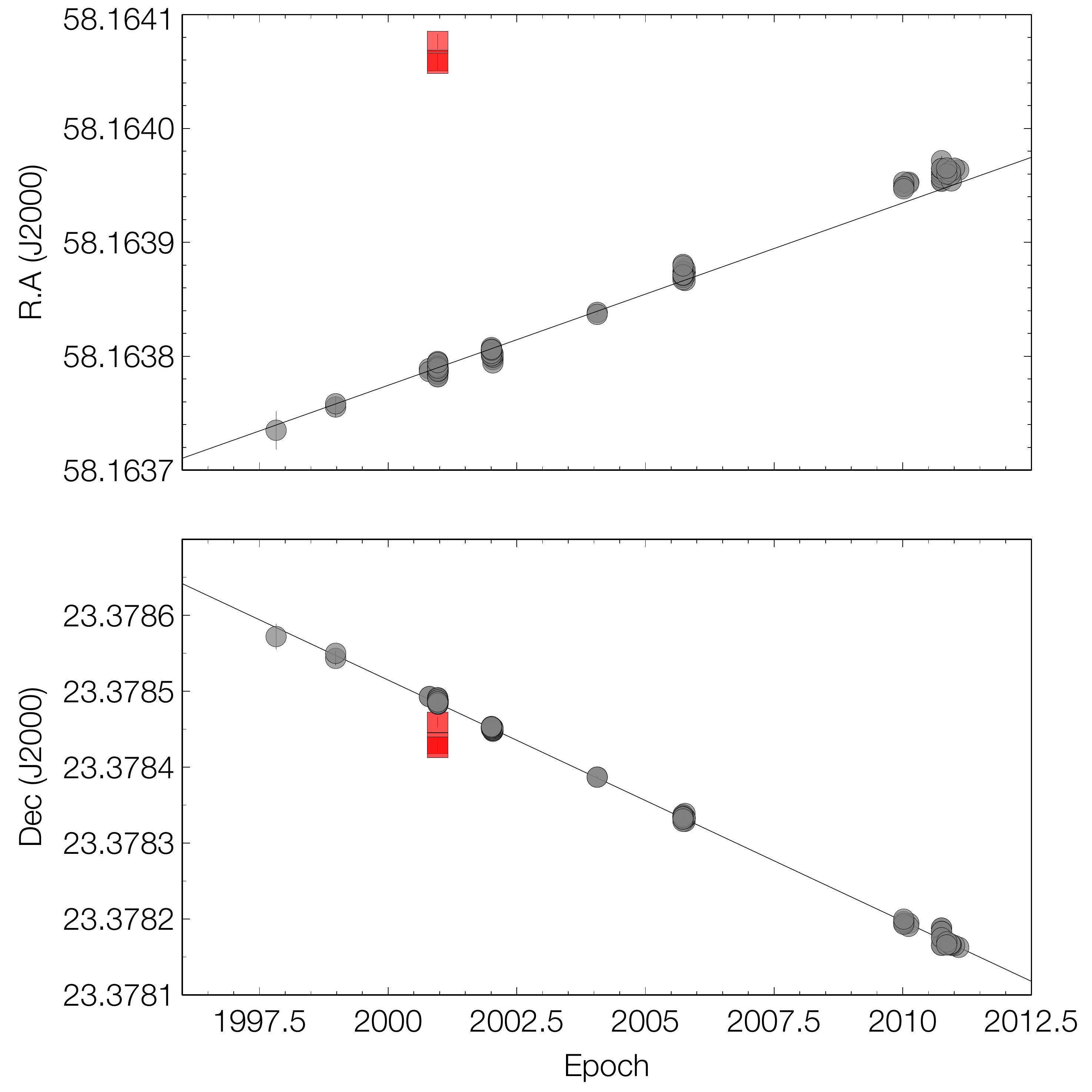}
      \caption{Example of proper motion fit in Right Ascension (upper panel)
	and Declination (lower panel). Red squares correspond to measurements
	rejected by the outlier filtering procedure. For most measurements the uncertainties are smaller than the symbol. A total of 93 individual
	exposures (out of 96 in total) were used for this source. See also Fig.~\ref{nearbyT}.}
      \label{sample_pm}
   \end{figure}

A prototype of empirical DCR correction is under development. It needs further
testing, and has been turned off for the present study. Systematic errors due to
DCR are nevertheless expected to be small:
\begin{itemize}
\item the vast majority (95\%) of the observations used in this study were
obtained at airmass $<$1.4. The resulting absolute DCR offsets for B stars can
add up to 30~mas in the $V$-band, 10~mas in the $I$-band and $<$1~mas in the
$H$ and $K$-bands under typical ambient conditions \citep{2002PASP..114.1070S}.
It goes down to 8, 7, 2 and $\ll$1~mas, respectively, for solar type dwarfs,
which corresponds to the high mass end of our sensitivity limit, and rises again to $\approx$25~mas in $V$ and 5~mas in $I$ for late M-dwarfs.
Relative offsets within the field-of-view of our instruments are expected to be even smaller.
\item the vast majority of the observations were obtained in the red or
near-infrared part of the spectrum, where the amplitude of the DCR is smaller.
\item several instruments used in this study are equipped with an atmospheric
dispersion compensator ({\it Subaru, CTIO/Mosaic2} and {\it KPNO/Mosaic1})
\item finally, for many sources, the effect of DCR on the proper motion fit is
averaged over the large number of individual measurements (see Fig.~\ref{sample_pm}).
\end{itemize}

\subsection{Charge Transfer Inefficiency}
Radiation damage of the CCD detectors can locally alter their charge tranfer efficiency (CTE), producing deformed PSF and affecting the source extraction accuracy. While this effect is important in the space environment, it is expected to be negligible at the level of accuracy of our study in ground based instruments. The complexity of the charge transfer inefficiency (CTI) effects and the large number of CCD instruments used in this study prevent us from attempting a systematic calibration. We nevertheless note that the dithering strategy used in the CCD observations is expected to average out the CTI effects on relative astrometry. Additionally, recent studies demonstrated that even a very low level of background strongly mitigates the CTI effects by filling the traps \citep{2012MNRAS.419.2995P}. The sky background in ground based CCD observations is several orders of magnitude higher than in space, and is expected to result in negligible CTI induced distortion of stellar images. Finally, a large fraction of the observations presented in this study was obtained with near-infrared detectors which are unaffected by CTI. We checked for a dependence of the residuals of the PSF fit and the astrometric solution with the signal-to-noise ratio and distance to the amplifier, but found no systematic distorsion or offset following the behavior expected for CTI effects. For the rest of the analysis, we consider the CTI effects to be negligible.

\subsection{Computing proper motions \label{ppm}}
\label{chap:propmot}
After the second iteration of the global astrometric calibration is completed,
{\sc SCAMP} performs another cross-matching of all detections, including those
that were rejected at the previous step.
{\sc SCAMP}'s cross-matching algorithm matches in priority detections found
in two or more {\em successive} exposures. We facilitate the cross-identification of
moving sources by feeding {\sc SCAMP} with exposures ordered by instrument and
by observation date. We adopt a cross-matching radius of 3\arcsec, which defines
the maximum  proper motion detectable in our study: $\approx 30$\arcmin/hr
for the highest exposure rate found in our sample (one every 10~s).

{\sc SCAMP} computes proper motions by doing a linear fit (in the weighted
$\chi^2$ sense) to source positions as a function of observation dates. No
attempt is made to include the effect of trigonometric parallaxes in the fit;
annual parallaxes would be poorly constrained for most sources because of
observation dates spanning a too short period of time each year.
It is not unusual for the position of a source in a given exposure to deviate
strongly from the linear trend with time expected from our model. Visual checks
indicate that this happens most
frequently because of a cosmetics problem (contamination by an electronic
glitch, a cosmic ray hit, a fringing pattern or an optical halo) or some
deblending issue. To detect and filter out outliers, {\sc SCAMP} applies a
specific procedure to sources with more than two valid epochs and enduring a
``bad'' fit, i.e. with a reduced $\chi^2$ above 6.
The procedure consists of removing from the fit the one detection that
decreases the most the reduced $\chi^2$, and iterate until it is less than 6
or a maximum of 20\% of points have been removed
(or 2 points if less than 10 points remain). With three detections, the
the pair that corresponds to the lowest proper motion is selected.
The filtering procedure is triggered on less than 5\% of sources.

\begin{figure}
   \centering
   \includegraphics[width=0.48\textwidth]{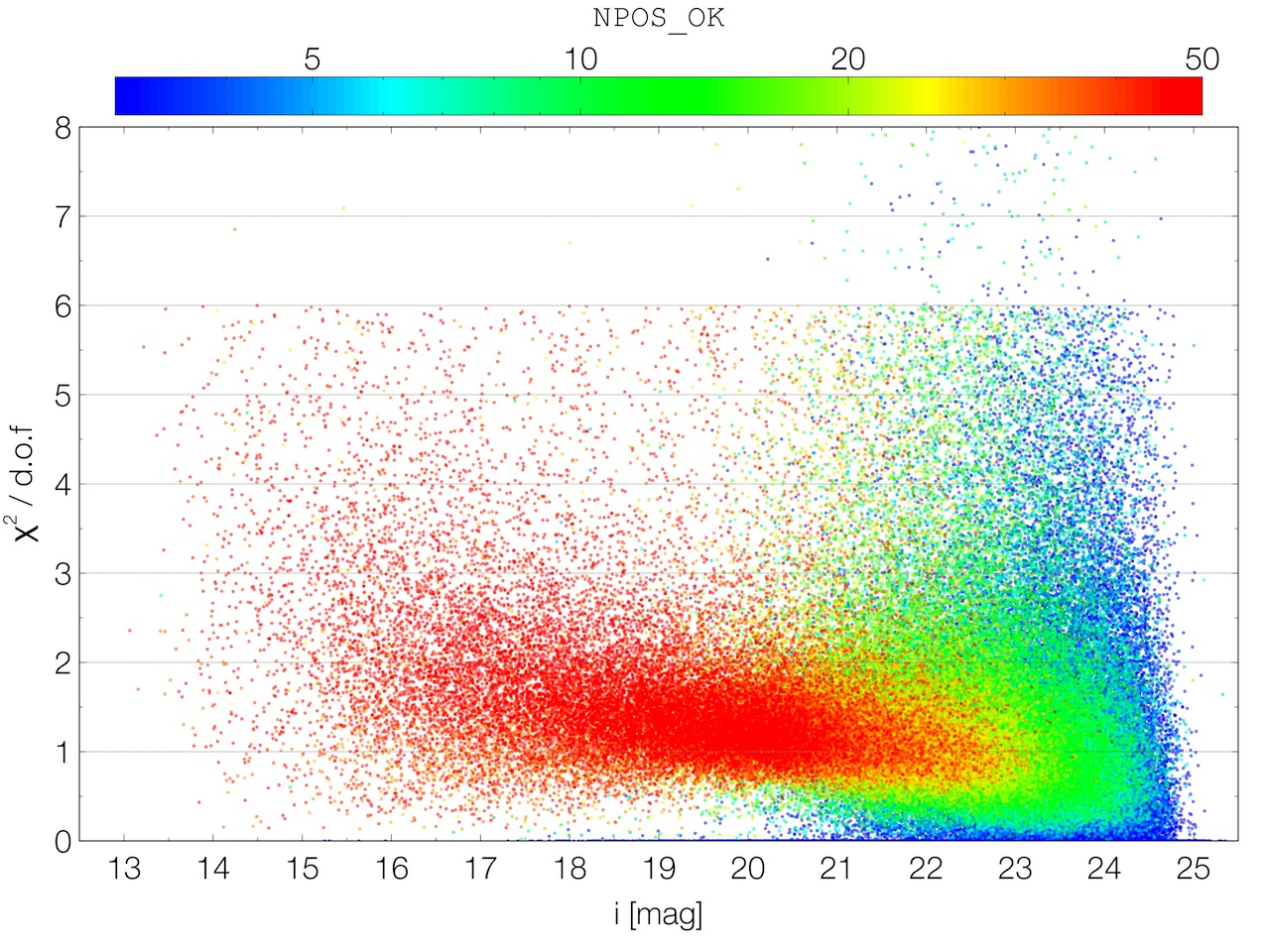}
      \caption{Reduced $\chi^{2}$ of the proper motion fit as a function of
	the {\it Megacam} $i$-band magnitude (when available), with the number of measurements {\tt NPOS\_OK}
	indicated by color. The cut-off at $\chi{^2}/{\rm d.o.f}$=6
	corresponds to the outlier rejection threshold (see text). For clarity, only 10\% of the catalogue is represented. }
      \label{fig:chi2_imag}
   \end{figure}

Figure~\ref{fig:chi2_imag} shows the distribution of the reduced $\chi^{2}$ as
a function of magnitude and the number of measurements used in the proper motion
fit. The reduced $\chi^{2}$s have values close to one over a large range of
magnitude, a hint that the estimated measurements are robust and their
uncertainties are reasonably well estimated.

\subsection{Anchoring to an absolute reference frame}
\label{chap:anchor}
\begin{figure*}
   \centering
   \includegraphics[width=0.95\textwidth]{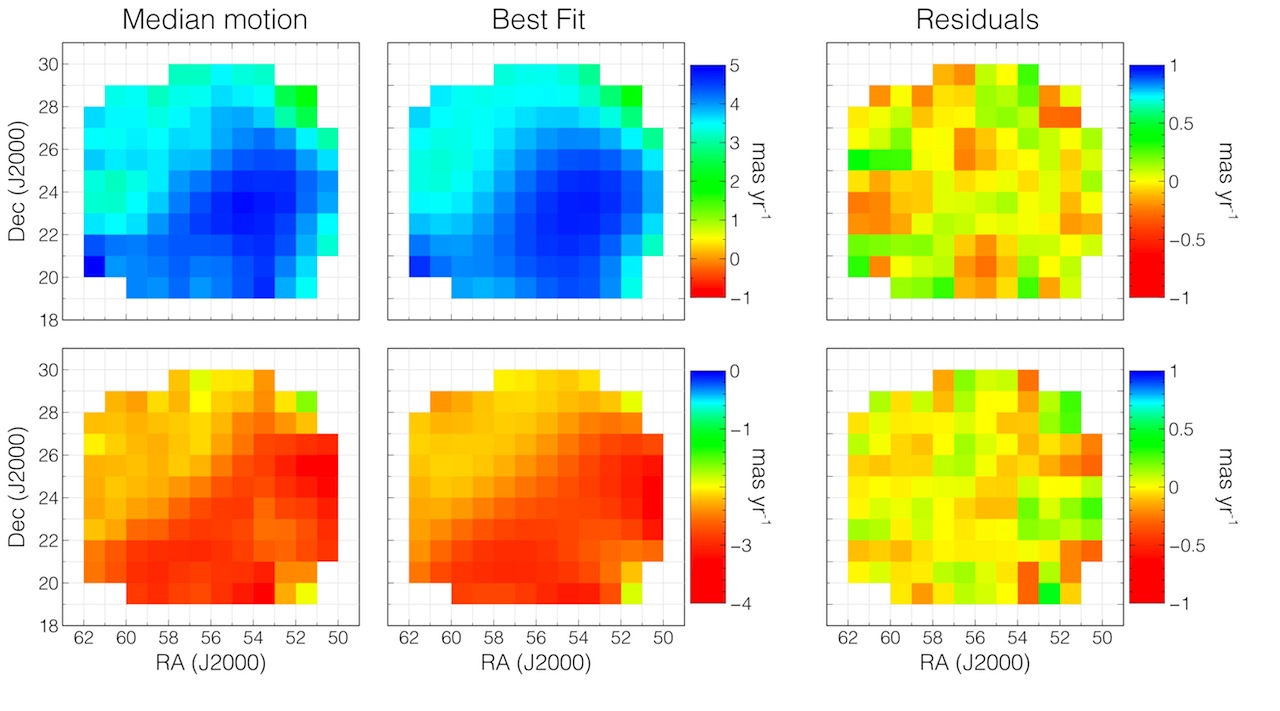}
      \caption{Distribution of median proper motion of extragalactic sources in right ascension $\mu_{\alpha} \cos \delta$ (upper panels) and declination $\mu_{\delta}$ (lower panels). A gradient oriented towards the galactic plane  is clearly visible. We adjust a 4$^{\rm th}$ order polynomial surface (middle panels). The residuals of the fit are shown in the right panels. }
         \label{fig:residuals_pm}
   \end{figure*}

The proper motions computed by {\sc SCAMP} are not explicitly tied to an
absolute reference system such as the ICRS (International Celestial Reference
System). Linking our measurements to the ICRS can be made by comparing to the
{\it Hipparcos} catalog. Unfortunately most stars from the {\it Hipparcos} and
{\it Tycho} catalogs present in the Pleiades field are saturated in our data.
The resulting large uncertainties on the corresponding position and proper
motion measurements prevent us from deriving an accurate offset to the  ICRS.
Nevertheless we can tie our kinematic measurements very closely to the ICRS
by computing the offset required to cancel out the apparent proper motion of extragalactic objects. The vast
majority of galaxies detected in our sample are resolved under sub-arcsecond
seeing conditions, and can therefore be easily and securely identified based on
their {\tt SPREAD\_MODEL} value (Fig.~\ref{fig:spreadmodel}). Even though the
astrometric precision is considerably worse for extended objects compared to
point sources, the large number of resolved extragalactic sources allows a
statistically meaningful and accurate calculation of the offset to the ICRS.
We select all sources with a  {\tt SPREAD\_MODEL} indicative of an extended
object that display a  proper motion less than 30~mas yr$^{-1}$ in both
R.A and Dec ($\approx$379\,000 sources), and compute their median proper
motion within boxes of 1\degr$\times$1\degr. Figure~\ref{fig:residuals_pm} shows
the spatial distribution of median apparent proper motions $\mu_{\alpha} \cos \delta$ and $\mu_{\delta}$. A gradient
pointing towards the galactic plane is clearly seen in both components, which we
interpret as the contribution of galactic stars to the overall astrometric
solution derived by our algorithm. To correct for these systematic motions, we
fit a 4$^{\rm th}$ order polynomial surface to apparent motions in our
``extragalactic'' dataset and use it to correct all individual measurements. The
residuals after correction are $<$0.2~mas yr$^{-1}$, and we conservatively add 0.2~mas yr$^{-1}$ quadratically to the final estimated uncertainty on the absolute proper motion.

\begin{figure}
   \centering
   \includegraphics[width=0.45\textwidth]{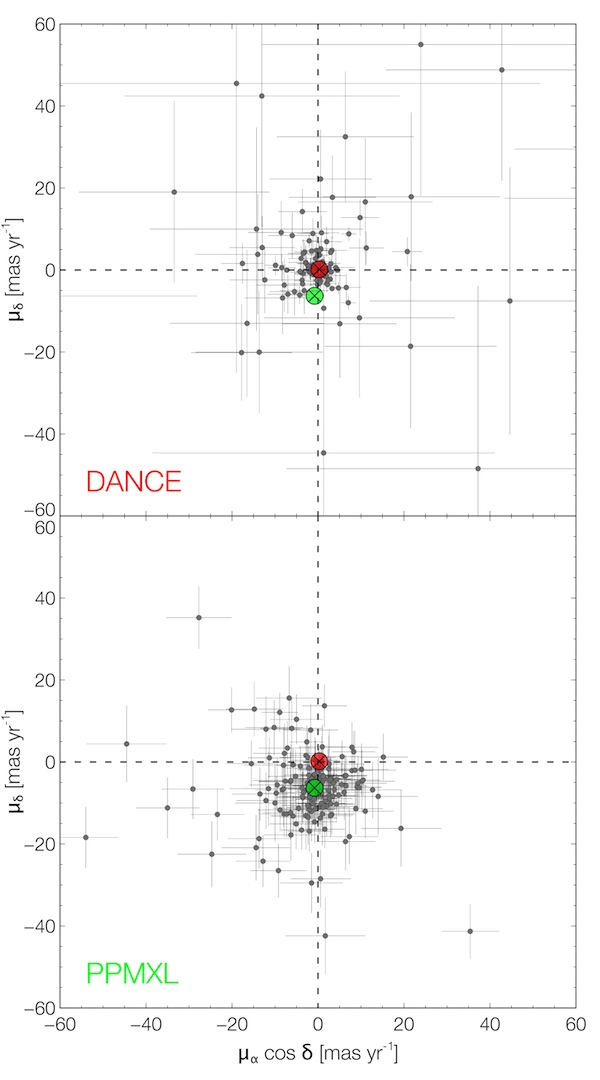}
      \caption{Upper panel: vector point diagram for the 126 quasars from the Million Quasars Catalogue with a counterpart in the DANCe catalogue. Lower panel: vector point diagram for the 164 quasars with a counterpart in the PPMXL catalogue. In both diagrams, the median value  of the DANCe (red) and PPMXL (green) is represented by a large color dot.}
         \label{fig:qso}
   \end{figure}

As a sanity check, we then compare the proper motions of the 126 quasars from the Million Quasars ({\it MILLIQUAS}) Catalogue (v.3.0) with a counterpart in our catalogue. Figure~\ref{fig:qso} shows the vector point diagram obtained. As expected, the median proper motion is very close to zero  ($\mu_{\alpha} \cos \delta$, $\mu_{\delta}$)= (0.29, 0.17)~mas~yr$^{-1}$.

\section{Photometric solution}
A global photometric solution is computed for each photometric instrument.
A photometric instrument is defined here as a set of instruments sharing a
unique photometric behavior. In the case of our study, we chose to define one
instrument per combination of telescope plus detector plus filter set. For
example, although they are very similar (see Fig.~\ref{filters}) the CFH12K
$i$-band photometric calibration will be treated independently from that of the
MegaCam $i$-band. This choice is made to minimize the effect of color terms
between the various physical instruments. Similarly to astrometry, the
photometric solution is computed through weighted $\chi^2$ minimization of the
quadratic sum of magnitude differences between overlapping detections from pairs
of exposures observed with the same photometric instrument. Color terms are
ignored and the only free parameters are the magnitude zero-points. Wherever
applicable, photometrically calibrated fields act as ``anchors'' in the final
solution. No zero-point correction is applied to isolated fields. No attempt is
currently made to derive illumination corrections for the various instruments:
a uniform zero-point is computed for each exposure. 
Finally, the absolute zero-point calibration provided by the observatories or derived using standard fields obtained the same night is accurate to 0.01 to 0.05~mag for images
obtained under clear or photometric conditions.

\section{Limitations}

\subsection{Accuracy and sources of errors}
The absolute astrometric accuracy is largely limited by the precision of the anchoring onto the extragalactic reference frame, and is described in the previous section. The residuals add up to a maximum of 0.2~mas yr$^{-1}$ rms over the $\approx$10\degr\, of the survey. The overall ``internal'' (or relative, image-to-image) accuracy of the calibration is largely limited by the distorsion corrections residuals and the variable anisokinetism related to atmospheric turbulences. We also identify a number of sources of errors that can affect the proper motion measurements:
\begin{itemize}
\item cosmic rays and bad pixels can make chance coincidences and therefore add noise to the astrometric solution. Their contribution can be greatly minimized by: i) using the most up-to-date bad pixel masks for each instrument, ii) cleaning non-overlapping images using Laplacian edge detection \citep{2001PASP..113.1420V}, iii) filtering abnormal measurements (in particular based on the  {\tt SPREAD\_MODEL} of the sources, see Fig.~\ref{fig:spreadmodel}), iv) rejecting outliers in the proper motion fit (see section~\ref{ppm}). 
\item artefacts produced by saturated stars (such as deformed point spread functions, streaks and bleeding due to pixel overflows) will seriously compromise the astrometric solution. This effect is minimized by carefully setting the saturation levels in {\sc SExtractor} input parameter files. We note that {\sc SExtractor} PSF fitting module is capable of adjusting a PSF to the non-saturated pixel of a source, extending the dynamic range of our study above the saturation and non-linearity regime of the instruments used in this survey. The corresponding astrometry, although less precise, is nevertheless often good enough to derive relatively accurate proper motion, as illustrated in Fig.~\ref{fig:dance_vs_tycho}.
\item extragalactic sources and nebulosities are often extended and their centroid position can be wavelength dependent. They can also surround point-like sources. The corresponding chromatic shift between overlapping images obtained in different filters can compromise the proper motion measurements, but also adds noise to the instrumental distortion measurement, and hence to the astrometric solution. The latter effect is minimized by adjusting carefully {\sc SExtractor's} parameters and by iteratively selecting only clean point-like sources for the astrometric registration, as described above. In particular, the  {\tt SPREAD\_MODEL} filtering is expected to efficiently reject extended sources. Finally, these chromatic shifts are expected to be stochastic in orientation and amplitude, and their effect on the global solution should average out.
\item unresolved multiple systems and visual binaries: the orbital motion of true multiple systems and the chromatic shift of blended pairs made of stars of different colors are an additional source of error that cannot be corrected for. Visual multiple systems resolved in a set of images and unresolved in another (e.g when the seeing or sensitivity are different) can also produce mismatches and errors. They usually result in large reduced-$\chi^{2}$ value.
\item differential chromatic refraction errors, as discussed in section~\ref{dcr}
\item parallax motion: at an average distance of $\approx$120~pc \citep{2009A&A...497..209V}, the maximum amplitude of the parallax motion of Pleiades members is of the order of $\approx$8~mas yr$^{-1}$. Our observations were obtained over yearly periods of approximately 4 months, and Pleiades members possibly display significant parallax motion. This effect is even larger for nearby stars. Our observational strategy and the archival observations were not designed to measure parallaxes, and the multi-epoch images are not suited for a good parallax determination, which also adds noise to the astrometric solution and proper motion fit. We nevertheless verify that most of these sources (and in particular the Pleiades members) are rejected by the 1-$\sigma$ clipping and their contribution to the astrometric solution is expected to be negligible.
\item atmospheric turbulence, as discussed in section~\ref{chap:astrom_error}. Using several consecutive observations allows to further average out this effect, and is an additional justification for using individual frames rather than stacked mosaics. 
\item proper motions themselves: although stars exhibiting large deviations caused by proper motions are clipped, moderate motions ($\approx$5-10 mas.yr$^{-1}$) may degrade the astrometric solution, affecting particularly the distortion patterns derived for the earliest and the latest runs in the observation time range. A new iterative procedure that recomputes the solution after correcting the positions of stars for the derived motions is under development in SCAMP, but it was not judged robust enough in its present state to be applied to this study.
\end{itemize}

\begin{figure*}
   \centering
   \includegraphics[width=0.95\textwidth]{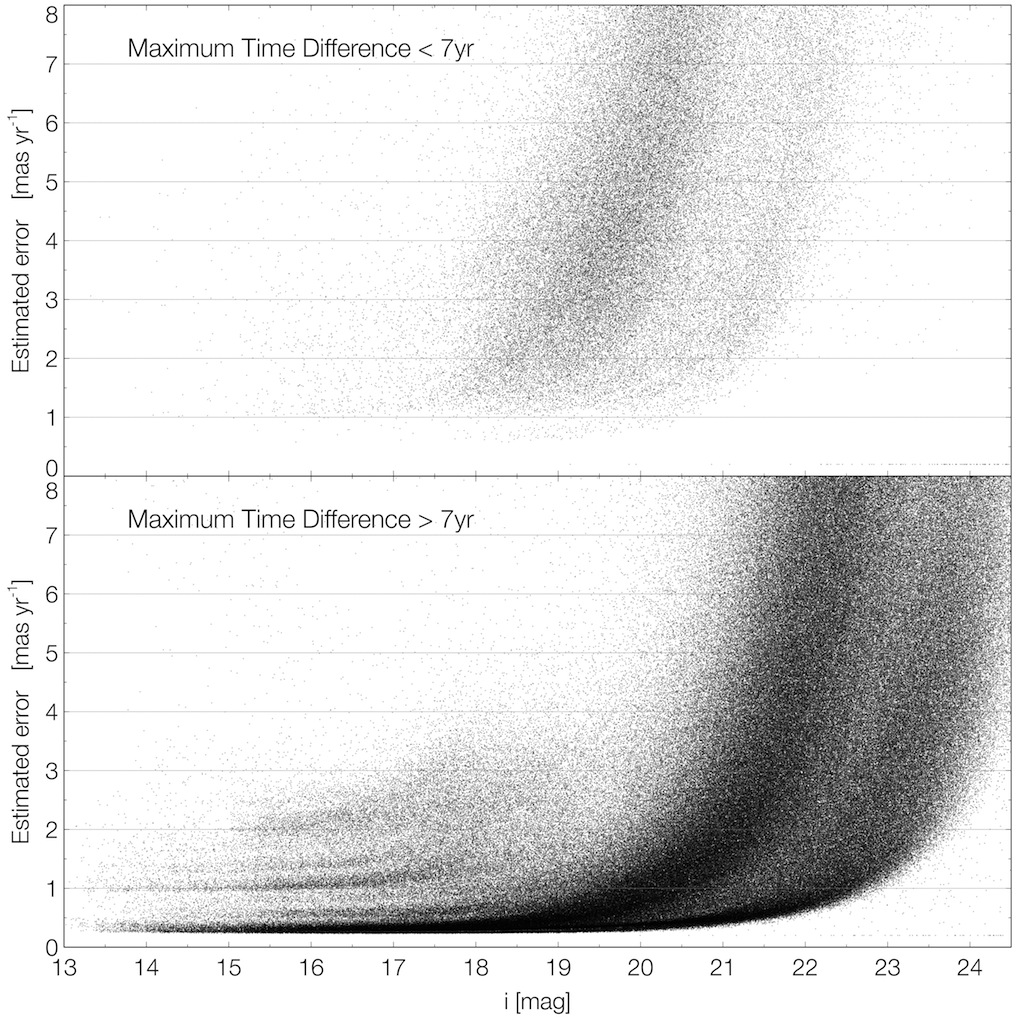}
      \caption{Estimated error of the absolute proper motion measurements as a function of MegaCam $i$-band magnitude. Upper panel: for sources with a maximum time difference less than 7~yr. Lower panel: for sources with a maximum time difference greater than 7~yr.}
         \label{fig:error}
   \end{figure*}

Fig.~\ref{fig:error} shows the estimated error of the absolute proper motion fit\footnote{computed as the quadratic sum of the RA and Dec components, and the 0.2~mas.yr$^{-1}$ residual related to the anchoring onto the ICRS}  as a function of the {\it MegaCam} $i$-band magnitude and the maximum time difference used for the fit. As expected, the estimated error is tightly correlated to the maximum time difference and to the luminosity. 

\section{Comparison with other astrometric catalogues \label{comparison}}
In the following, we compare our measurements to various astrometric databases found in the litterature and check the consistency of our results. Several of these catalogues are not tied to any absolute reference frame (e.g UCAC4, UKIDSS), and a direct comparison with the DANCe catalogue (anchored on background galaxies) is therefore not strictly correct. We nevertheless note that the difference can in general be approximated to a simple offset.

\subsection{Tycho}
\begin{figure*}
   \centering
   \includegraphics[width=0.95\textwidth]{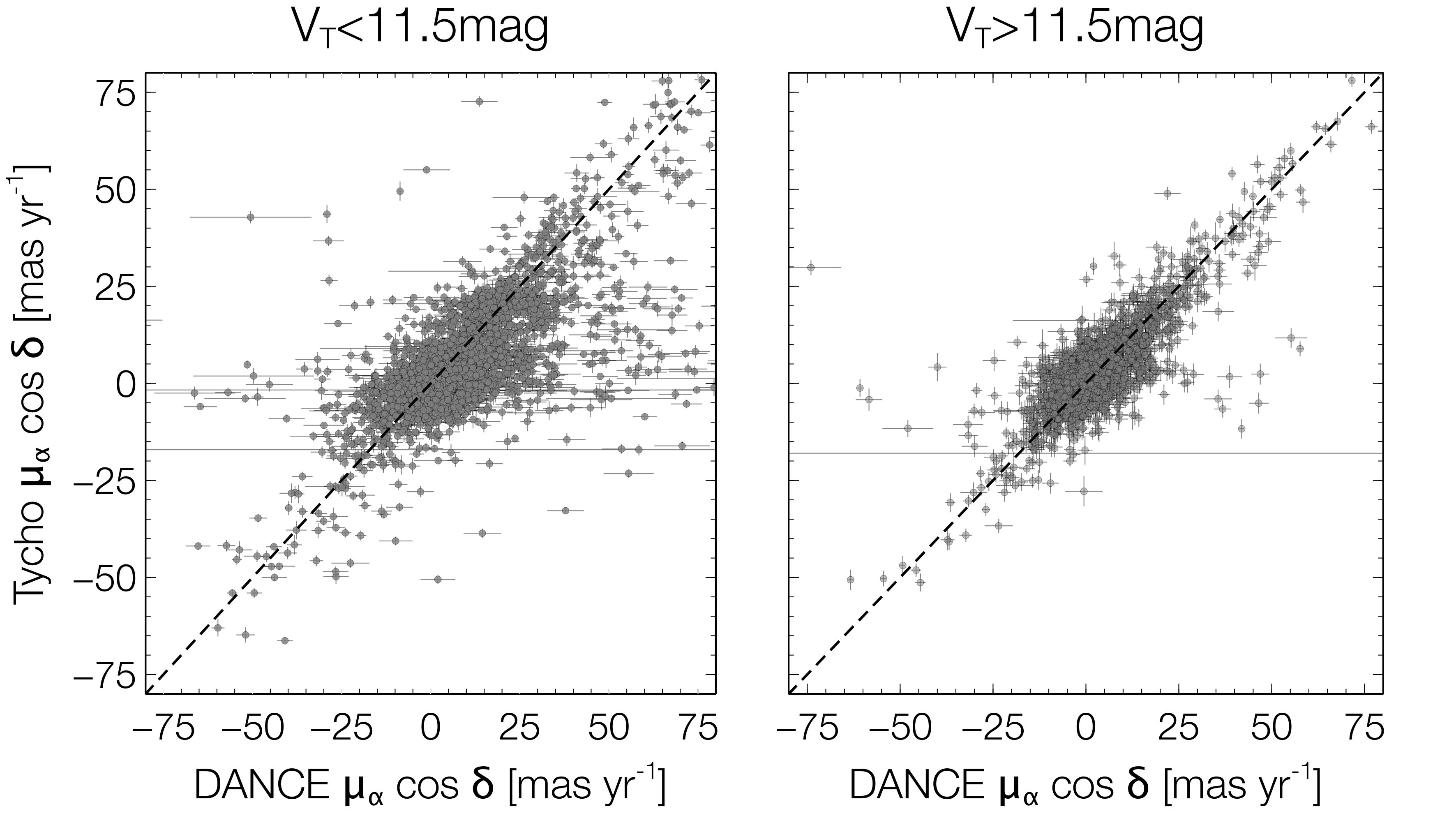}
      \caption{Proper motion in RA for the DANCe (x-axis) and Tycho (y-axis) catalogues. The error bars represent the estimated error in the case of the DANCe measurements, and the reported uncertainty in the case of the Tycho measurement. The left panel corresponds to sources with V$_{\rm T}<$11.5mag, and the right panel to sources with V$_{\rm T}>$11.5mag. A dashed line corresponding to a linear relation is represented to guide the eyes. A similar distribution is found in declination.}
         \label{fig:dance_vs_tycho}
   \end{figure*}

The Tycho catalogue \citep{2000A&A...355L..27H} provides proper motion measurements precise to about 2.5~mas yr$^{-1}$ and derived from a comparison of the ESA {\it Hipparcos} satellite measurements with the Astrographic Catalogue and 143 other ground-based astrometric catalogues. These catalogues are unfortunately limited to $V_{T}\lesssim$12.5~mag, very close to the saturation or non-linear regime of the datasets used to build the DANCe catalogue. With this limitation in mind, we compare the results obtained for the 3665 common sources. The agreement is good within the large uncertainties, as shown in Fig.~\ref{fig:dance_vs_tycho}. As expected, the difference between the Tycho and DANCe measurement displays a clear dependance on the luminosity, fainter sources being in general in better agreement than bright sources.  

\subsection{UCAC4 \label{comp_ucac4}}
\begin{figure*}
   \centering
   \includegraphics[width=0.95\textwidth]{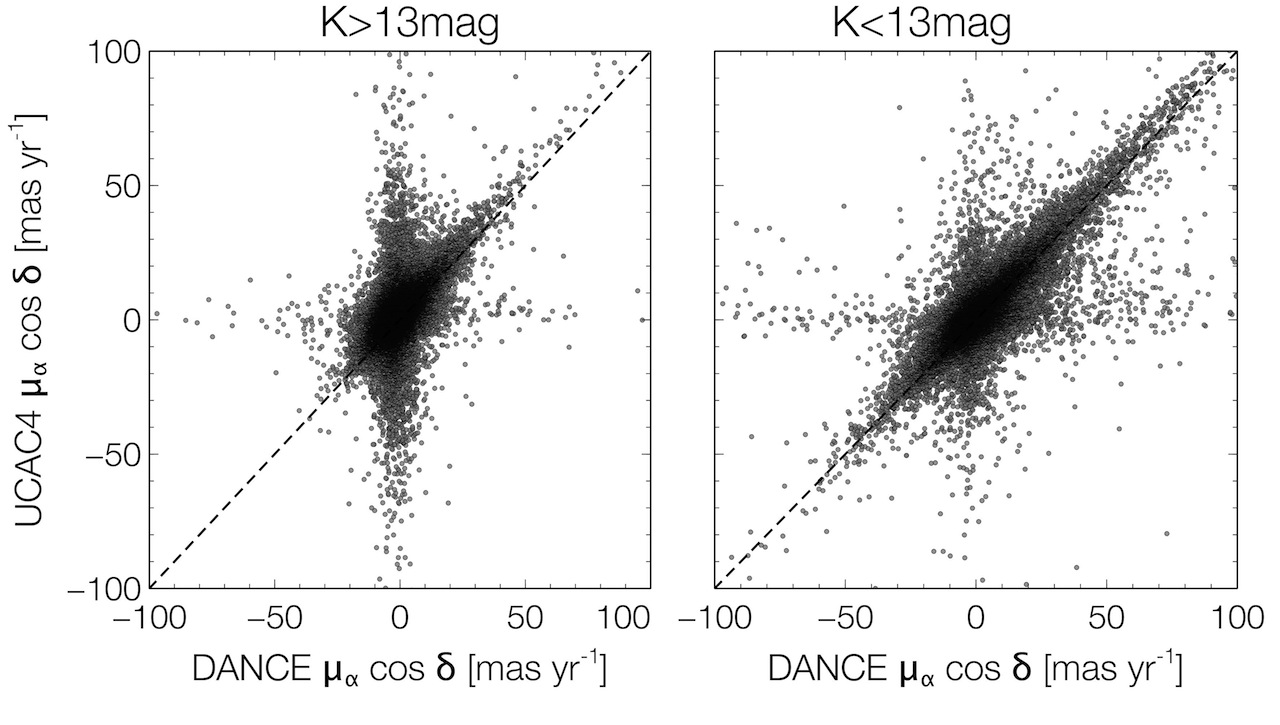}
      \caption{Proper motion in RA for the DANCe (x-axis) and UCAC4 (y-axis) catalogues. Only a random subsample corresponding to 10\% of the total number of matches is represented for clarity. The left panel corresponds to sources fainter than K=13mag, and the right panel to sources brighter than K=13mag. A dashed line corresponding to a linear relation is represented to guide the eyes.  A similar distribution is found in declination.}
         \label{fig:dance_vs_ucac4}
   \end{figure*}

The Fourth US Naval Observatory CCD Astrograph Catalog \citep[UCAC4, ][]{2010AJ....139.2184Z} provides astrometry, photometry and proper motion measurements over the entire sky and covering the luminosity range between 8$\lesssim$R$\lesssim$16~mag. Uncertainties are typically of the order of 1--10~mas yr$^{-1}$, depending on magnitude and observing history. Figure~\ref{fig:dance_vs_ucac4} shows a comparison of the proper motion measurements in RA for the UCAC4 and DANCe surveys, as a function of 2MASS Ks-band luminosity. Three major groups of sources can be identified: 
\begin{enumerate}
\item ``vertical outliers'' are sources with close-to-zero motion in DANCe but significant motion in UCAC4
\item ``horizontal outliers'' are sources with close-to-zero motion in UCAC4 but significant motion in DANCe.
\item sources with motions in good agreement in both catalogues within the typical uncertainties and to a constant offset
\end{enumerate}

We find that both outlier groups are clearly related to the luminosity of the sources: the ``vertical outliers'' are in general among the faintest sources, where UCAC4 is less accurate and contains more errors. The ``horizontal outliers'' are in general bright sources, and we interpret them as erroneous or inaccurate measurements due to saturation and/or non-linearity of the DANCe datasets. As expected, the distribution of sources is asymetric, with significantly more sources along the direction of the solar antapex ($\mu_{\alpha}$cos$\delta>$0 and $\mu_{\delta}<$0), which coincidently also corresponds to the Pleiades cluster's mean motion direction. 

\subsection{PPMXL \label{comp_ppmxl}}
\begin{figure*}
   \centering
   \includegraphics[width=0.95\textwidth]{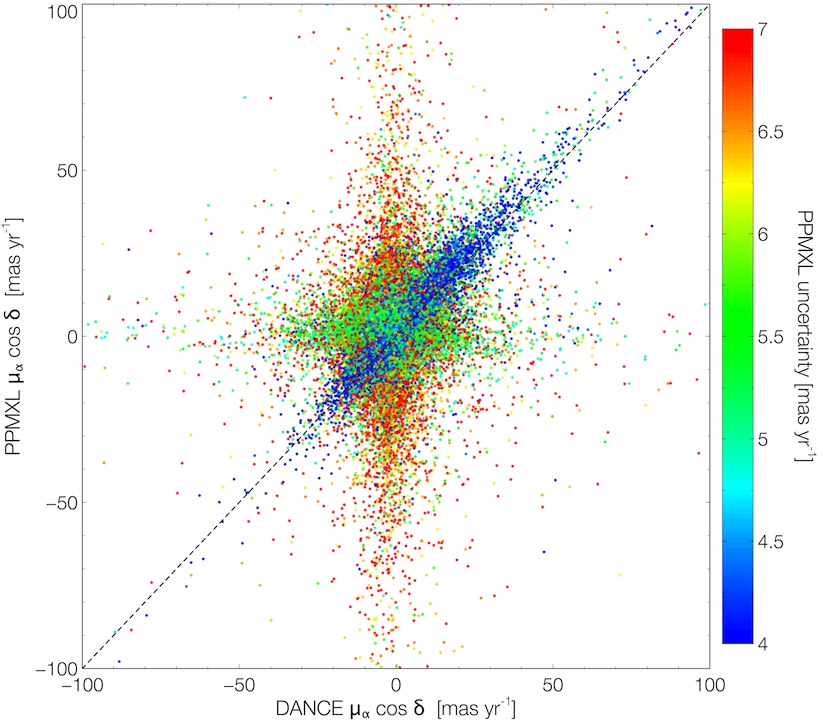}
      \caption{Proper motion in RA for the DANCe (x-axis) and PPMXL (y-axis) catalogues.The color scale represents the PPMXL uncertainty. A similar distribution is found in declination. A dashed line corresponding to a linear relation is represented to guide the eyes.}
         \label{fig:dance_vs_ppmxl}
   \end{figure*}

\citet{2010AJ....139.2440R} derived improved mean positions and proper motions on the ICRS system by combining USNO-B1.0 and 2MASS astrometry. The catalog is complete from the brightest stars up to about $V\approx$20~mag over the entire sky. Typical individual errors of the proper motions range between  4--10~mas yr$^{-1}$. Figure~\ref{fig:dance_vs_ppmxl} compares the proper motion measurements in RA and Dec for the PPMXL and DANCe surveys as a function of the PPMXL uncertainty. The same three groups of sources described in Section~\ref{comp_ucac4} can be seen:
\begin{enumerate}
\item ``vertical outliers'' are sources with close-to-zero motion in DANCe but significant motion in PPMXL. They generally have one or all of the following properties: a) they have the fewest \emph{No} number of measurements in PPXML, b) they are among the faintest sources and c) they have a flag \emph{fl}=1 in the PPMXL catalogue indicative of a problematic fit. On the other hand, They seem to have a reasonable maximum time difference in the DANCe survey, which is in general associated to more reliable proper motion measurements.
\item ``horizontal outliers'', with close-to-zero motion in PPMXL but significant motion in DANCe. These sources generally have the smallest maximum time difference and only a few individual measurements in the DANCe catalogue, suggesting that the DANCe proper motion measurements are less reliable. 
\item sources with motions in good agreement in both catalogues within the typical uncertainties
\end{enumerate}

We also note that in general, the 2 outlier populations are made of the faintest sources, for which the astrometric precision is expected to be lower. 
 
Figure~\ref{fig:dance_vs_ppmxl} also shows an offset between the DANCe and PPMXL measurements, especially obvious in declination. Figure~\ref{fig:qso} suggests that the PPMXL proper motion measurements are indeed offsetted with respect to the ICRS, as quasars from the MILLIQUAS catalogue do not have an average zero motion in the PPMXL catalogue. We note that a similar offset is also reported in the SPM4 catalogue \citep[see Fig.~8 of][]{2011AJ....142...15G}.

\subsection{UKIDSS DR9}
\begin{figure}
   \centering
   \includegraphics[width=0.45\textwidth]{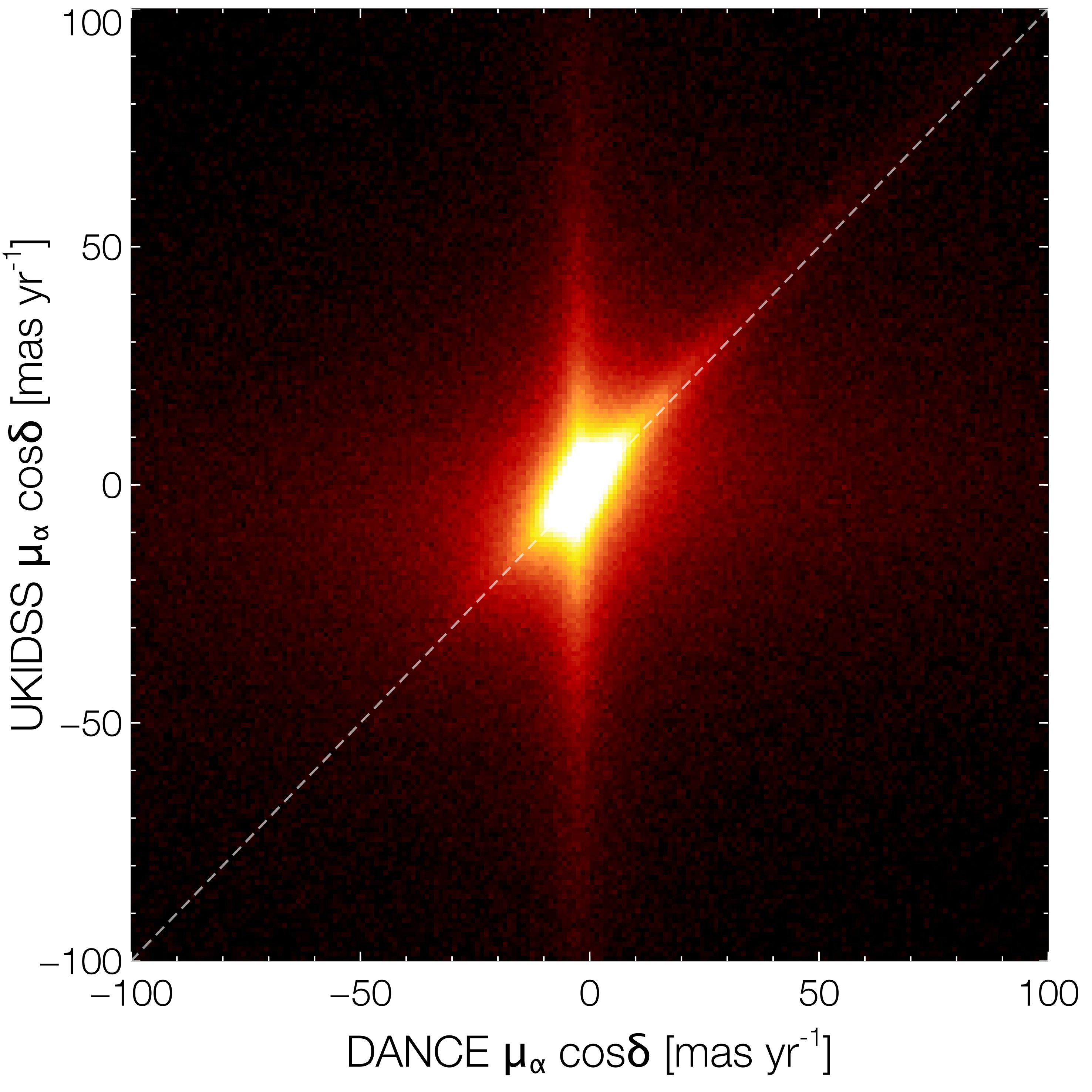}
      \caption{Density map of the proper motion in RA for the DANCe (x-axis) and UKIDSS (y-axis) catalogues. A dashed line corresponding to a linear relation is represented to guide the eyes. The same behaviour is found in declination.}
         \label{fig:dance_vs_ukidss}
   \end{figure}

The Pleiades cluster was observed as part of the UKIDSS Galactic Cluster Survey. \citet{2012MNRAS.422.1495L} recently presented a photometric and astrometric study based on the corresponding catalogue, which includes proper motion measurements based on the multi-epoch UKIDSS observations. The proper motion measurements given in the UKIDSS DR9 catalogue provide a useful comparison as the DANCe survey includes all the UKIDSS individual images. Figure~\ref{fig:dance_vs_ukidss} compares the proper motion measurements in RA for the UKIDSS and DANCe catalogues. 
 Three major groups of sources appear clearly:
\begin{enumerate}
\item ``vertical outliers'' are sources with close-to-zero motion in DANCe but significant motion in UKIDSS. In UKIDSS they generally have a) the fewest measurements (\emph{nFrames} attribute); b) the highest star/galaxy classifier value, indicative of high probability to be an extended extragalactic sources; and c) the largest time difference in DANCe
\item sources with motions in good agreement in both catalogues within the typical uncertainties and to a constant offset corresponding to the offset of the UKIDSS measurements to the ICRS.
\item a diffuse group of sources in poor agreement. Most of these sources are detected in the UKIDSS dataset only, and in general have a small maximum time difference resulting in larger uncertainties in the proper motion fit in both catalogues. The lose correlation and dispersion of this group are consistent with the typical uncertainties ($>$25~mas yr$^{-1}$) of the corresponding measurements in both catalogues.
\end{enumerate}

As the DANCe measurements for the vertical outliers correspond to the most probable extragalactic sources (which are supposed to have no detectable motion, in agreement with the DANCe measurements) and to the measurements with the largest DANCe time baseline (hence more robust in general), we are confident that the DANCe measurements of the vertical outliers are more reliable than their UKIDSS counterparts. We interpret the inconsistency for the vertical outliers' population as a greater sensitivity of the UKIDSS proper motion fit to deviant individual astrometric measurements. The small number of measurements used in the proper motion fit ($\leq$6) together with the lack of rejection in the UKIDSS proper motion fit \citep{2012MNRAS.422.1495L} make it much more sensitive to the presence of deviant and high leverage points and translates into large numbers of errors. The presence of corrupted frames in the UKIDSS DR9 release (discarded by our quality assurance but not by the UKIDSS quality assurance, M. Read private comm.) probably also result in a number of problematic measurements. By using the individual UKIDSS images rather than stacked UKIDSS mosaics, and by including a robust regression algorithm for the proper motion fit, our method is much less sensitive to erroneous individual measurements, as demonstrated by the lack of a clear ``horizontal outliers'' population.

\subsection{On the use of the DANCe and other astrometric catalogues}
Large catalogues necessarily contain errors and problems. The comparison of the DANCe measurements to other astrometric catalogues calls for a number of important warnings about their use:
\begin{itemize}
\item some proper motion measurements are more reliable than others, and the uncertainty does not always reflect the reliability. Parameters useful to evaluate the reliability of an individual proper motion measurement include in particular (but not exhaustively): the number of astrometric measurements used for the fit, the maximum time baseline, the reduced-$\chi^{2}$ of the fit
\item measurements errors, unknown systematics and problematic measurements present in any large scale astrometric catalogue most likely always affect the completeness of studies based on their proper motion measurements and should be carefully discussed
\end{itemize}

While a universal rule to assess the quality of a given measurement cannot be given, we have found that the following proper motion measurements should be considered with caution:
\begin{itemize}
\item sources close to or above the saturation or linearity limit of the instruments (in the case of the current dataset, $i\approx$13~mag), or
\item sources with small numbers of measurements used for the proper motion fit (\verb|NPOS_OK| attribute)
\item sources with large reduced-$\chi^{2}$  (\verb|CHI2_ASTROM| attribute)
\end{itemize}

In general, and whenever possible, a visual inspection of the individual images is the most robust way to discard problematic measurements.

\section{Example of scientific applications}

The DANCe catalogue includes accurate photometry and astrometry for 6\,116\,907 unique sources, and proper motion measurements for 3\,577\,478 of them, and as such represents a unique opportunity to address various scientific problems. In the following, we give a few examples of direct applications making use of the catalogue.

\subsection{The Pleiades cluster}
\begin{figure}
   \centering
   \includegraphics[width=0.48\textwidth]{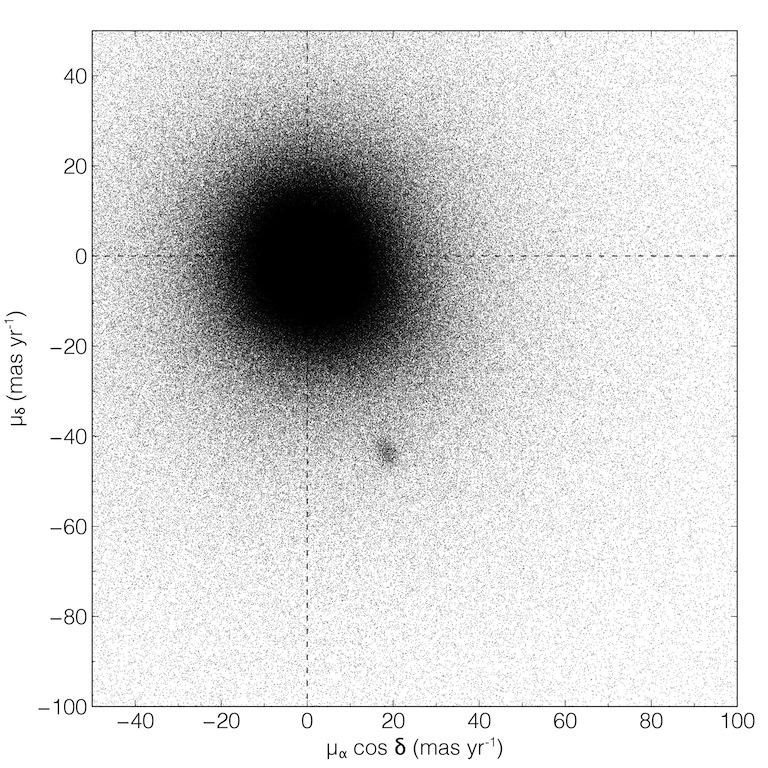}
      \caption{Vector point diagram of stellar motions obtained with the datasets described in this article. The Pleiades locus is visible in the lower right quadrant. Its slight elongation may be interpreted as a perspective effect related to the depth of the cluster.} 
         \label{vpd}
   \end{figure}

A detailed scientific analysis of the Pleiades cluster kinematics based on the DANCe catalogue will be presented in a future article (Bouy et al., 2012, in prep.). In this section, we give a brief and general overview of the results obtained. Figure~\ref{vpd} shows the vector point diagram of stellar motions obtained with the dataset described above. The group of co-moving cluster members appears clearly around (20,-40)~mas yr$^{-1}$.

\begin{figure}
   \centering
   \includegraphics[width=0.48\textwidth]{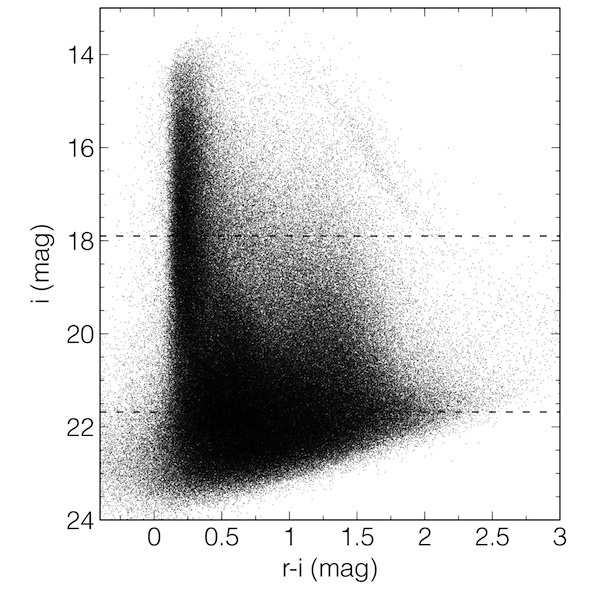}
      \caption{$i$ vs $r-i$ color magnitude diagram ({\it MegaCam}). Two horizontal lines represent the luminosity of Pleiades members with masses 0.072 and 0.030~M$_{\odot}$, according to the models of \citet{1998A&A...337..403B} and assuming of distance of 120~pc. The Pleiades sequence is visible.}
         \label{cmd}
   \end{figure}

The DANCe catalogue also offers a unique photometric database. In the case of the Pleiades dataset presented here, a total of 29 photometric instruments covering the spectral range between 0.37~$\mu$m (Sloan $u$-band) and 2.2~$\mu$m (UKIRT $K$-band). Figure~\ref{cmd} shows a $i$ vs $r-i$ color magnitude diagram using the photometry extracted from the MegaCam and UKIDSS images. The cluster sequence is also visible. 

\subsection{Solar system bodies}
\label{chap:asteroids}
\begin{figure}
   \centering
   \includegraphics[width=0.35\textwidth]{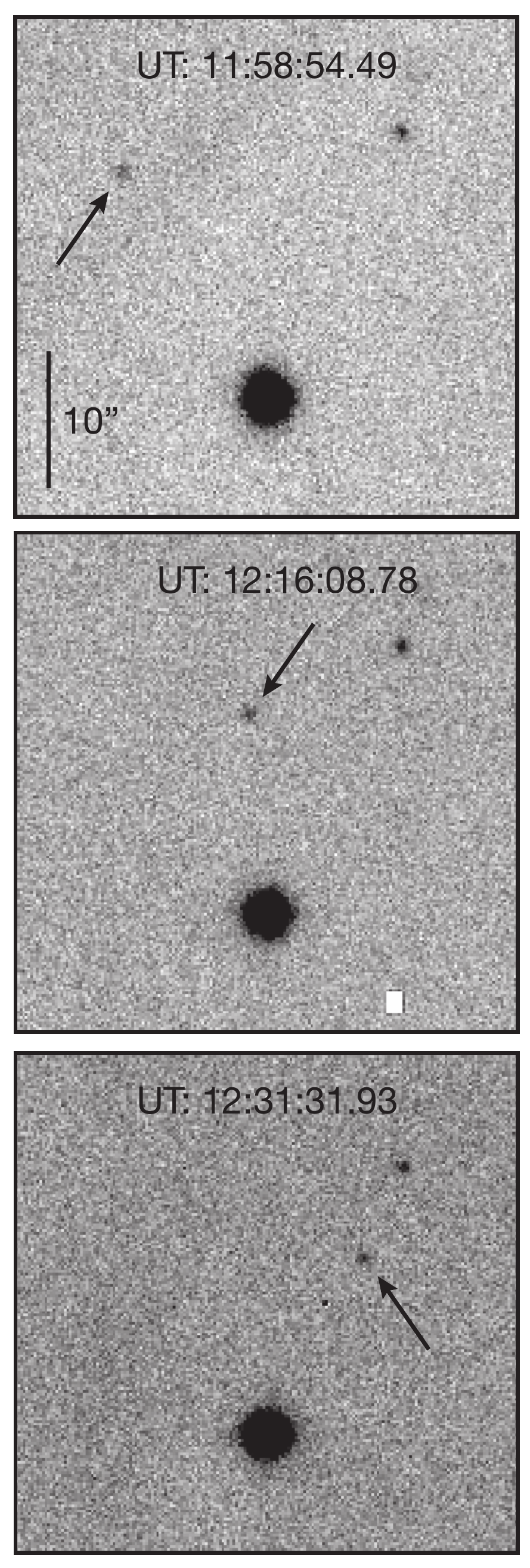}
      \caption{Successive \emph{CFHT MegaCam} images of a main belt asteroid of $g$=23.1~mag. The UT time of observations are indicated. These 3 images are an illustrative subset of 19 images used to measure the proper motion of this source.}
         \label{asterodance}
   \end{figure}

Solar system bodies have typical velocities in the range between $\approx$2\arcsec\,hr$^{-1}$ (trans-neptunian objects, hereafter TNO) and $\approx$20\arcsec\,hr$^{-1}$ (main belt asteroids, hereafter MBA). Most observations used in this study are made of several consecutive and dithered images of the same field. Fast moving sources such as solar system bodies can therefore be easily identified. Minor planets are expected to be extremely numerous in the direction of the Pleiades cluster, as it lies close to the ecliptic plane. A detailed analysis of the solar system bodies encountered in the DANCe dataset will be presented in a future article (Bouy et al. in prep) and we only give a brief overview of the capabilities of our algorithms for solar system studies. A basic selection of all sources with a proper motion greater of $\approx$20\arcsec\,hr$^{-1}$ gave 11404 candidate minor planets. A request on {\it SkyBot} \citep{2006ASPC..351..367B} indicates that only 2837 have a counterpart within a radius of 1\arcmin\,  in the database of known solar system bodies as of August 2012, all of them main belt asteroids. A visual inspection of $\approx$100 random candidates shows that $\lesssim$5\% are false detections due to artefacts (cosmic ray or ghost coincidence, false detection, etc...). After inspection and rejection of these artefacts, the astrometric and photometric measurements will be submitted to the IAU Minor Planet Center. The high precision astrometry (with a typical accuracy better than $\lesssim$10~mas on individual epochs) will be extremely valuable to refine the orbital solutions. The accurate photometry (with typical absolute accuracy $<$10\%, and relative accuracy better than $<<$1\%) will be useful to classify the bodies (based on their colors) and in some cases study their rotational periods and geometry. The depth of the datasets allows the discovery of very faint objects, as illustrated in Fig.~\ref{asterodance}, probing a largely incomplete asteroid size domain.

\subsection{Nearby ultracool dwarfs}
\begin{figure*}
   \centering
   \includegraphics[width=0.95\textwidth]{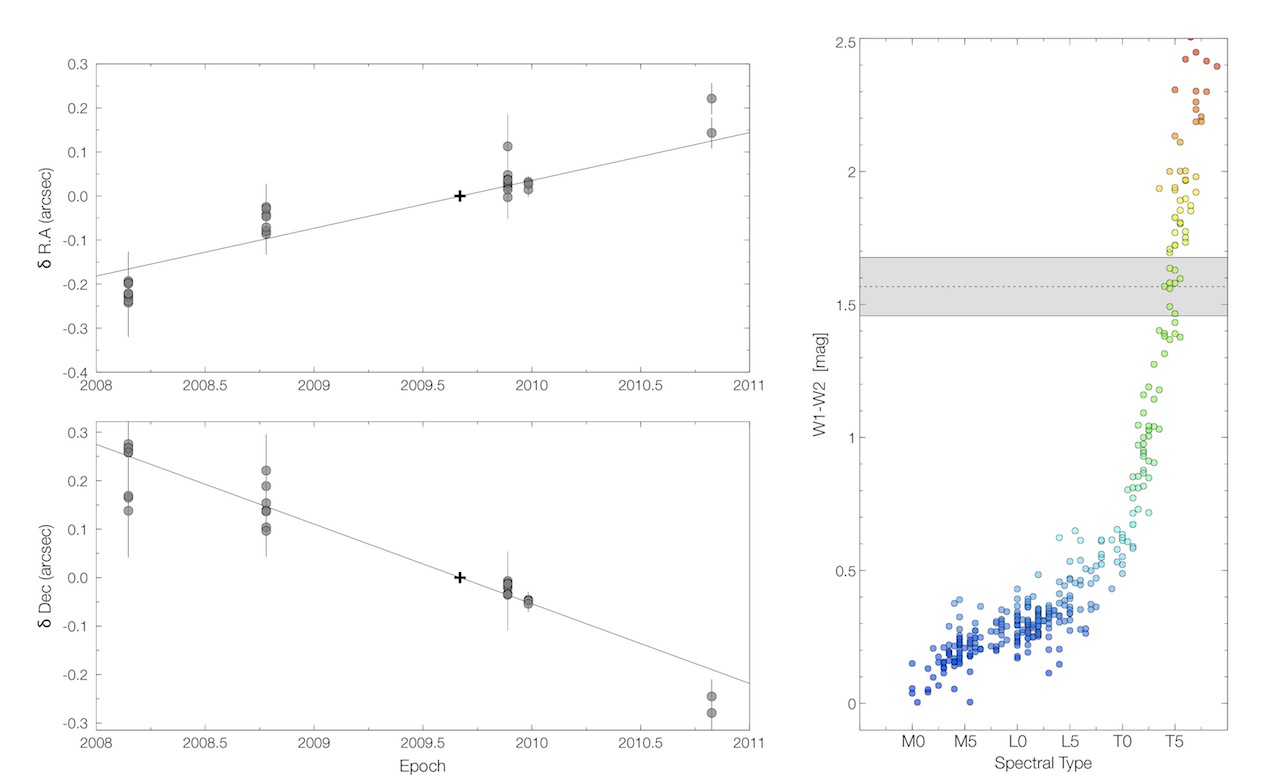}
      \caption{Relative motion in RA (upper left panel) and Dec (lower left panel) of a candidate T5 dwarf discovered in the survey. \emph{WISE} [W1]-[W2] colors of ultracool dwarfs from \citet{2011ApJS..197...19K}. The color of our new candidate is indicated with an horizontal line, and the uncertainty domain is represented by a light grey area.}
         \label{nearbyT}
   \end{figure*}

Nearby ultracool dwarfs can easily be identified in the DANCe catalogues as faint fast moving sources. Figure~\ref{nearbyT} shows an example of such object discovered in the present survey. A complete analysis will be presented in a future paper, and we here only present the basic properties of one particular source to illustrate the scientific case. The source must be relatively nearby as it moves at $\approx$200~mas yr$^{-1}$. It has a counterpart in the \emph{WISE} \citep{2012yCat.2311....0C} and its [W1]-[W2] color matches that of known T4$\sim$T5 ultracool dwarfs from \citet{2011ApJS..197...19K}. 

\subsection{Galactic dynamics}
\begin{figure}
   \centering
   \includegraphics[width=0.45\textwidth]{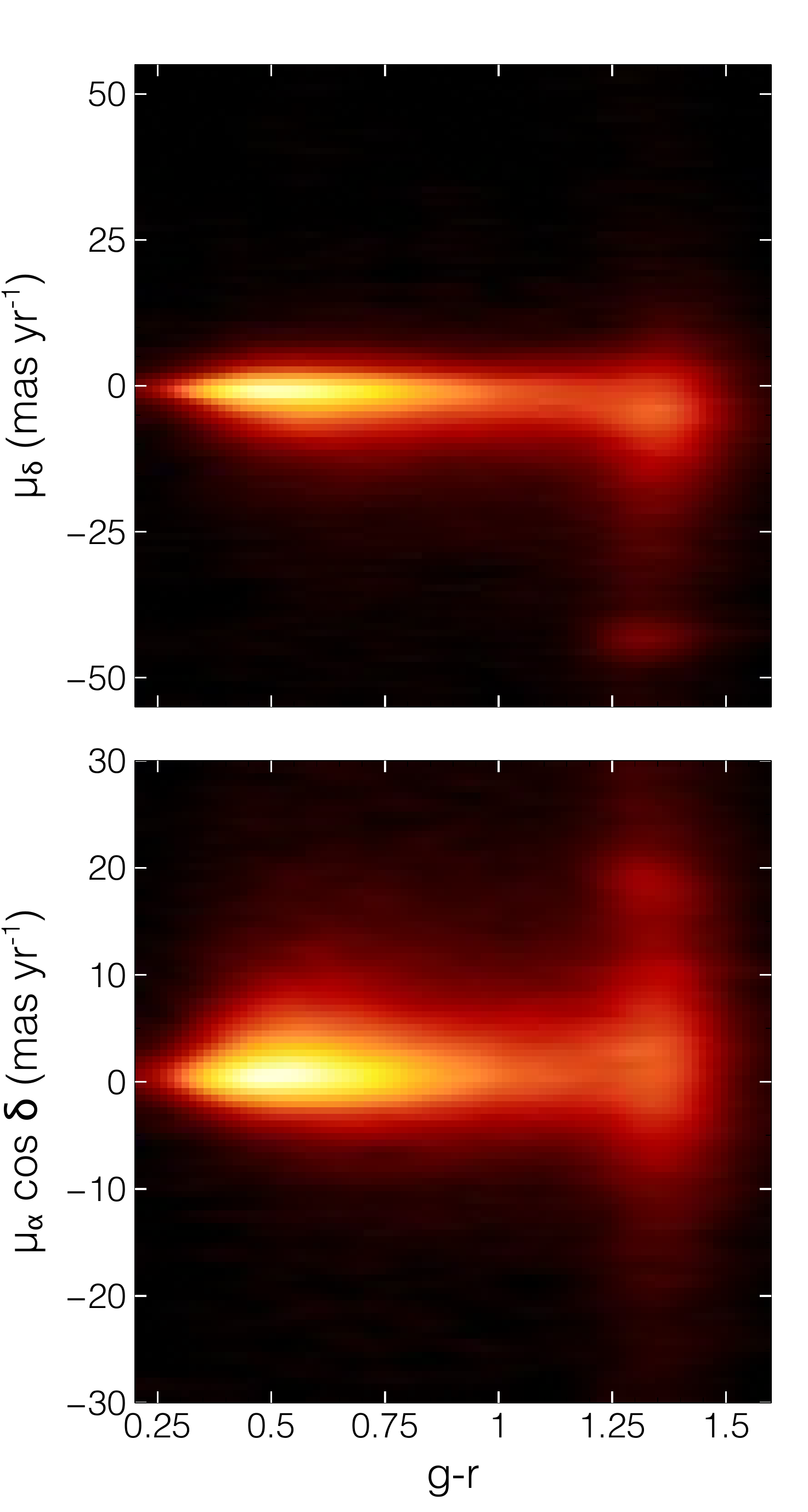}
      \caption{Density maps showing the distribution of proper motion in RA (upper panel) and Dec (lower panel) as a function of the $g-r$ color for galactic sources in the range 12$<g<$21~mag. }
         \label{fig:pm_gr_DANCe}
   \end{figure}

Accurate large scale photometric and astrometric surveys provide a unique opportunity to study the galactic stellar populations. In Fig.~\ref{fig:pm_gr_DANCe}, we show the distribution of motions in RA and Dec as a function of the $g-r$ color. For this figure, a subset of galactic sources was selected in the DANCe catalogue based on:
\begin{itemize}
\item the quality of the proper motion measurement, keeping sources with $g$ magnitude in the range 12--21~mag where the estimated uncertainties are better than $\lesssim$2~mas yr$^{-1}$ in average.
\item the ``stellarity'', rejecting all sources with a {\tt SPREAD\_MODEL} indicative of an extended source. Although it does not reject unresolved extragalactic sources,  the remaining extragalactic contamination on the luminosity range mentioned above should be small enough for the simple purpose if this illustrative example
\end{itemize}
A gross bimodal structure in $g-r$ is clearly seen, reflecting the separation of the halo/thick-disk ($g-r\sim$0.5~mag) and the thin-disk ($g-r\sim$1.3~mag) populations. The Pleiades population is clearly seen as a small clump around (20,-40)~mas, yr$^{-1}$ on top of the general thin disk population. Such observations can provide very important constraints and input to the models of galactic populations.

\section{Conclusions and future prospects}
We have presented a set of tools capable of deriving high precision relative proper motions using large numbers (several thousands) of ground-based images originating from various instruments. We apply these tools on multi-epoch panchromatic datasets of the nearby Pleiades cluster, and compare to other astrometric catalogues. The results demonstrate our ability to derive accurate proper motion with an estimated accuracy better than 1~mas yr$^{-1}$ for sources as faint as $i$=22$\sim$23~mag, depending on the luminosity and observational history (time baseline, number and quality of the frames, as well as presence and number of reference extragalactic sources for the anchoring on the ICRS). 

The DANCe project will use this method to conduct a survey of the most nearby star forming regions and clusters. It aims at complementing the {\it Gaia} mission in the substellar regime, and in regions of high extinction. By taking advantage of the wide field surveys performed in the late 90's and early 2000's, it will provide high precision proper motion measurements for millions of stars in various nearby associations. In the future, the DANCe project will also take advantage of the growing number of wide and very wide field imagers which will equip various observatories. Large and all-sky surveys are also on-going (e.g Pann-STARRS,ESO-VST, ESO-VISTA) or foreseen (DES, LSST), ensuring a large flow of high quality images useable for high precision astrometry.

\begin{acknowledgements}
We are thankful to our anonymous referee for a thorough and constructive review which helped significantly improve this article.
H. Bouy is funded by the the Ram\'on y Cajal fellowship program number RYC-2009-04497. H. Bouy acknowledges funding and support of the Universit\'e Joseph Fourier 1, Grenoble, France. This research has been funded by Spanish grants AYA2012-38897-C02-01, AYA2010-21161-C02-02, CDS2006-00070 and PRICIT-S2009/ESP-1496. E. Moraux ackowledges funding from the Agence Nationale pour la Recherche program ANR 2010 JCJC 0501 1 ``DESC (Dynamical Evolution of Stellar Clusters)''. J. Bouvier acknowledges funding form the Agence Nationale pour la Recherche program ANR 2011 Blanc SIMI 5-6 020 01 (``Toupies''). J. Bouvier and E. Moraux acknowledge support from the Faculty of the European Space Astronomy Centre (ESAC). A. Bayo ackowledges funding and support from the Marie Curie Actions of the European Commission FP7-COFUND.
E. Bertin acknowledges partial funding of computer resources by the French Programme National de Cosmologie et Galaxies and CNRS-Fermilab contract \#367561.
We are grateful to Mike Read and Mike Irwin for their help and assistance with the INT and UKIRT archives, and to Mark Farris and James Beletic at Teledyne for looking into the pixel grid issue with the WFCam Hawaii-2 detectors.

Based on observations obtained with MegaPrime/MegaCam, a joint project of {\it CFHT} and CEA/DAPNIA, at the Canada-France-Hawaii Telescope (CFHT) which is operated by the National Research Council (NRC) of Canada, the Institut National des Science de l'Univers of the Centre National de la Recherche Scientifique (CNRS) of France, and the University of Hawaii. This paper makes use of data obtained from the Isaac Newton Group Archive which is maintained as part of the CASU Astronomical Data Centre at the Institute of Astronomy, Cambridge. The data was made publically available through the Isaac Newton Group's Wide Field Camera Survey Programme. The Isaac Newton Telescope is operated on the island of La Palma by the Isaac Newton Group in the Spanish Observatorio del Roque de los Muchachos of the Instituto de Astrofísica de Canarias. This research used the facilities of the Canadian Astronomy Data Centre operated by the National Research Council of Canada with the support of the Canadian Space Agency. This research draws upon data provided by C. Briceño as distributed by the NOAO Science Archive. NOAO is operated by the Association of Universities for Research in Astronomy (AURA) under cooperative agreement with the National Science Foundation. This publication makes use of data products from the Two Micron All Sky Survey, which is a joint project of the University of Massachusetts and the Infrared Processing and Analysis Center/California Institute of Technology, funded by the National Aeronautics and Space Administration and the National Science Foundation. This work is based in part on data obtained as part of the UKIRT Infrared Deep Sky Survey.  This research has made use of the VizieR and Aladin images and catalogue access tools and of the SIMBAD database, operated at CDS, Strasbourg, France. This research has made use of IMCCE's SkyBoT VO tool. This publication makes use of data products from the Wide-field Infrared Survey Explorer, which is a joint project of the University of California, Los Angeles, and the Jet Propulsion Laboratory/California Institute of Technology, funded by the National Aeronautics and Space Administration. This research has made use of the Million Quasars (MILLIQUAS) Catalog, Version 3.0 (9 September 2012). The MILLIQUAS Catalog made use of NASA/IPAC Extragalactic Database (NED) which is operated by the Jet Propulsion Laboratory, California Institute of Technology, under contract with the National Aeronautics and Space Administration, and also made use of data obtained from the Chandra Source Catalog, provided by the Chandra X-ray Center (CXC) as part of the Chandra Data Archive.

\end{acknowledgements}

\bibliographystyle{aa}

\end{document}